\documentclass[]{pasj01}
\draft

\begin{document} 
\Received{}
\Accepted{}

\title{ 
New method for black-hole spin measurement based on flux variation from an infalling gas ring}

\author{Kotaro \textsc{Moriyama}\altaffilmark{}%
\thanks{}}
\altaffiltext{}{Department of Astronomy, Kyoto University, Kitashirakawa, Oiwake-Cho, Sakyo-ku, Kyoto 606-8502}
\email{moriyama@kusastro.kyoto-u.ac.jp}

\author{Shin \textsc{Mineshige}\altaffilmark{}}



\maketitle

\begin{abstract}

We propose a new method for black hole spin measurement. 
In this method, we consider a gas blob or ring falling onto a black hole from the marginally stable orbit, 
keeping its initial orbital angular momentum. 
We calculate the gas motion and photon trajectories in the Kerr space-time and, 
assuming that the gas blob or ring emits monochromatic radiation,
carefully examine how it is observed by a distant observer. 
The light curve of the orbiting gas blob is composed of many peaks because of 
periodic enhancement of the flux due to the gravitational lensing and beaming effects. 
Further, the intensity of each peak first gradually increases with time due to the focusing effect around the photon circular orbit 
and then rapidly decreases due to the gravitational redshift, as the gas blob approaches the event horizon.
The light curve of the gas ring is equivalent to a superposition of those of the blobs 
with various initial orbital phases, and so it is continuous and with no peaks.
The flux first gradually increases and then rapidly decays, as in the blob model.
The flux variation timescale depends on the black hole spin and is independent from the inclination angle, 
while time averaged frequency shift have dependences of both effects.
We can thus, in principle, determine spin and inclination angle from observations.
The observational implications and future issues are briefly discussed.
\end{abstract}

\section{Introduction}

Observational proof of the black hole space-time is one of the most outstanding issues in physics and astrophysics. 
This can lead to a proof of the existence of the event horizon and to a critical observational 
test of the general relativity theory proposed 100 years ago. 
It is known in general relativity that black hole space-time is uniquely described by a black hole mass, $M$, and a spin parameter, $a$ 
(other than charges, which are never important in the astrophysical context). 
Masses can be relatively easily estimated by observing the motions of stars or gas 
 (Shahbaz et al. 1999; Ghez et al. 2005; Orosz et al. 2011).
The observed targets may not be necessarily close to the black hole, as long as
they are bound to its gravitational potential.
Black hole spins are, on the other hand, not easy to measure, 
since they only influence the space-time near to the black hole 
so that full considerations of general relativistic effects should be necessary.\
 
Until now, several methods have been proposed for the spin measurements. 
There are three major methods: 
(i) continuum spectrum method (Hanawa 1989; Li et al. 2005; McClintock et al. 2011; McClintock et al. 2014), 
(ii) line spectrum method (Kojima 1991; Laor 1991; Tanaka et al. 1995; Duro et al. 2011; Steiner et al. 2011; Reynolds 2014), 
and (iii) quasi-periodic oscillation (QPO) method (Kato 2001; Strohmayer 2001; Rezzolla et al. 2003; Remillard 2005; Kato et al. 2008). 

We wish to note, however, that all of these methods have uncertainties and/or are 
based on the critical assumptions that are not always easy to prove. 
The continuum spectrum method (i), for example, critically depends on the inclination angle. 
It also has an uncertainty in the spectral hardening factor in the emission from an accretion disk. 
The line spectrum method (ii) is sensitive to the emissivity profile assumed for the illuminated flux, as well as the iron ionization fraction as a function of radius. 
Further it is not obvious how to separate line spectra from continuum ones.
Moreover, it is assumed in both of methods (i) and (ii) are constructed on the assumption that the inner edge of the disk is
at the radius of marginally stable orbit, but this remains as a controversial issue (see Shidatsu et al. 2014).
In the QPO method (iii) no widely accepted model is available for the origins of quasi periodic oscillations. 
Further, several different modes of QPOs are known and it is difficult to identify. 

We should also point that spins estimated by different methods are sometimes inconsistent. 
For example, the spin of GRO 1655-40 estimated by methods (i) and (iii) are  $0.65<a/M<0.75$
and $0.31<a/M<0.42$, respectively (Shafee et al. 2006; Kato et al. 2008; McClintock et al. 2011). 
The spin of GRS 1915+105 is estimated to be $a/M>0.98$ [method (i)], 
$0.54<a/M<0.58$ or $0.97<a/M<0.99$ [(ii)], and $a/M<0.44$ [(iii)], respectively
(Kato et al. 2008; Blum et al. 2009; McClintock et al. 2011). 
In short, all the existing methods are far from being complete.
Therefore, it is required to construct new and independent methods for spin measurement 
to check other methods. 

That is, we consider the gas blob or ring falling into the black hole from the marginally
stable orbit (or the inner edge of an accretion disk) with a finite angular momentum 
and carefully examine how the gas blob or ring is observed by a distant observer. 
Radiation from the gas is known to undergo five relativistic effects 
which are summarized in table 1 (see Karas et al. 1992; $\rm{Dov\check{c}iak} $ et al. 2004a). 
We calculate the gas motion and photon trajectories in the Kerr space-time by general relativistic
 ray-tracing method to seek for good observational indicators of the black hole spin. 
Note that our approach is distinct from that by $\rm{Dov\check{c}iak} $ et al. (2004b) 
who were concerned with radiation from an accretion disk whose inner radius can be set below the marginally stable orbit, 
whereas we consider the gas blob or ring which is separated from an accretion disk and falls from the radius to the black hole. 

The plan of this paper is as follows: 
in section 2, we introduce the basic equations to describe gas motions and ray trajectories, 
explain our models for the gas blob and ring, and describe methods of calculations. 
In section 3, we show numerical results and explain the physics underlying the key features of the light variations.
On the basis of these simulation results we propose in section 4
our method for determining a spin $a$ and inclination angle $i$.
Section 5 is devoted to discussion on the distinctive features of our method,
its observational implications, and future issues.

	\begin{table}
	\tbl{Five relativistic effects on radiation emitted by gas near a black hole.}{%
	\scalebox{1.2}[1.2]{ 
	\begin{tabular}{llll}\hline
	     & Relativistic effect                 &  Physical cause                                                    & Remark (reference)\\ \hline
		(I)  & Gravitational lensing         & Focusing by light bending                                         & (Karas et al. 1992)\\  
		&										      & towards a gravitating source									&							\\ \hline			
		(II) & Beaming                          & Energy boost by relativistic                                     & (Karas et al. 1992)\\
		&                                           & motion of particles                                                 & \\  \hline
		(III) & photon circular orbit       & Focusing around the                                              & key feature in the early stage\\
		&                                          &  photon circular orbit, $r_{\rm ph}$							& \\ \hline
		(IV) & Focusing by the              &    Focusing by frame dragging                               & effective when $a/M$ is high\\ 
		& frame dragging                      &    due to black hole spin                                      &\\ \hline
		(V) & Gravitational redshift       & Energy loss by light travel                              & key feature in the late stage \\
		&                                           & from deep potential well                                 &\\ \hline
	\end{tabular}}}\label{tab1}
	\begin{tabnote}
	\end{tabnote}
	\end{table}

\section{Models and methods of numerical calculations}

	In this section, we first introduce the basic equations (subsection 2.1).
	Next, we explain our models of an infalling gas blob (subsection 2.2) and ring (subsection 2.3). 
	Finally we describe methods of numerical calculations (subsection 2.4).
	
	\subsection{Geodesic equations of gas particles and photons}
	
		In Boyer-Lindquist coordinates, the black hole space-time is

		\noindent
		\begin{eqnarray} 
			ds^{2}=-\left( 1-\frac{2Mr}{\Sigma} \right) dt^{2} -\frac{4Mar\sin^{2}\theta}{\Sigma}dtd\phi + \frac{\Sigma}{\Delta}dr^{2}+\Sigma d\theta^{2} \nonumber \\
			+\left( r^{2}+a^{2}+\frac{2Ma^{2}r\sin^{2}\theta}{\Sigma} \right) \sin^{2}\theta d\phi^{2}, 
		\end{eqnarray}

		\noindent
		with 
		\begin{eqnarray}
		\left\{
		\begin{array}{l}
		\Delta  = r^{2}-2Mr+a^{2},\\
		\Sigma = r^{2}+a^{2}\cos^{2}\theta ,
		\end{array}
		\right.
		\end{eqnarray}
		
		\noindent
		where we shall use ``geometrized units", in which the gravitational constant, $G$, and the speed of light, $c$, are set to be unity.  
		The spin parameter of the black hole is $a=J/M$, where $J$ is the angular momentum of the black hole.

		Around the black hole, motions of a particle (with mass of $\mu$) and photon (with no mass) 
		are determined by the geodesic equations (Bardeen et al. 1972): 
		
		\noindent
		\begin{eqnarray}
			\displaystyle \Sigma \frac{dt}{d\lambda}        &=& -a(aE\sin^{2}\theta -L)+\frac{(r^{2}+a^{2})T}{\Delta}, \nonumber \\
			\displaystyle \Sigma \frac{dr}{d\lambda}        &=& \pm \sqrt{V_{r}}, \nonumber \\
			\displaystyle \Sigma \frac{d\theta}{d\lambda} &=& \pm \sqrt{V_{\theta}}, \nonumber \\
			\displaystyle \Sigma \frac{d\phi}{d\lambda}   &=& -\left( aE-\frac{L}{\sin^{2}\theta} \right) +\frac{aT}{\Delta}, 
		\end{eqnarray}
		\noindent
		with
		
		\noindent
		\begin{eqnarray}
			E &=&-p_{t}=\rm{const}, \nonumber \\
			L&=& p_{\phi}=\rm{const}, \nonumber  \\
			\displaystyle Q &=& p_{\theta}^{2}+\cos^{2}\theta \left[a^{2}(\mu^{2}-p_{t}^{2})+\frac{p_{\phi}}{\sin^{2}\theta} \right] =\rm{const}, \nonumber \\
			T &=& E(r^{2}+a^{2})-La, \nonumber \\
			V_{r} &=& T^{2}-\Delta[\mu^{2}r^{2}+(L-aE)^{2}+Q],  \nonumber \\	
			\displaystyle V_{\theta} &=& Q-\cos^{2}\theta \left[ a^{2}(\mu^{2}-E^{2})+\frac{L^{2}}{\sin^{2}\theta}\right ],
		\end{eqnarray}

		\noindent
		where $p_{\rm{\mu}}$ is the 4-momentum of a test particle, and $\lambda$ is related to 
		the proper time of the test particle, $\tau$, by $\lambda = \tau /\mu$. 
		Note that $\lambda$ is the affine parameter in the case $\mu\rightarrow 0$.
		In addition, $E$, $L$, and $Q$ are the energy, the angular momentum, and the Carter constant of a test particle, respectively 
		(Carter 1968).
		
		As for a photon trajectory, we set $\mu=0$ in equations (3) and (4).

	\subsection{Gas blob model}

		We postulate the situation that a part of the innermost region of an accretion disk is stripped off, 
		thus forming a gas blob, and that it starts to fall onto the black hole, keeping its original angular momentum.  
		We assume that the gas blob has the following properties: 
		\begin{enumerate}
		\item It has a spherical shape and its characteristic radius is $R_{\rm blob}$, which has no time dependence.
		\item We neglect self-gravity of the blob. 
		\item The gas blob emits radiation with monochromatic frequency, $\nu_{0}$, in its inertial frame.
		This assumption is valid even in the case of multicolorwavelength radiation, such as the blackbody radiation, 
		if the frequency width of the spectrum is sufficiently small.
		Its emissivity decreases with an increase of a distance from a center of the blob, obeying the Gaussian function. 
		\item We neglect self absorption within the blob.
		\item The initial position of the center of the gas blob is at $(r, \phi )= (0.98r_{\rm{ms}}, 0)$. 
		Here $r_{\rm{ms}}$ is the radius of the marginally stable orbit. 
		\item The motion of the center of the blob follows the one particle orbit given by equations (3) and (4).
		The blob falls to the black hole on the equatorial plane, keeping the constant energy, $E_{\rm ms}$, and angular momentum, $L_{\rm ms}$, 
		where $E_{\rm{ms}}$ and $L_{\rm{ms}}$ are the energy and angular momentum of the particle rotating on the marginally stable orbit (figure 1).
		\end{enumerate}		
		
		By assumptions 1 -- 4, the emissivity, $j_{\rm{\nu}}$, in the inertial frame of the blob can be expressed as

		\noindent
		\begin{equation}
			j_{\rm{\nu}}(R)=j_{0}\exp\left[ -\left( \frac{R}{R_{\rm{blob}}} \right)^{2} \right] \delta (\nu -\nu_{0}),
		\end{equation}

		\noindent
		where $R$ is a distance from the center of the blob, and  
		 $R_{\rm{blob}}$, $j_{0}$, and $\nu_{0}$ are numerical constants.	
		
		\begin{figure}
		\begin{center}
		\includegraphics[width=10cm]{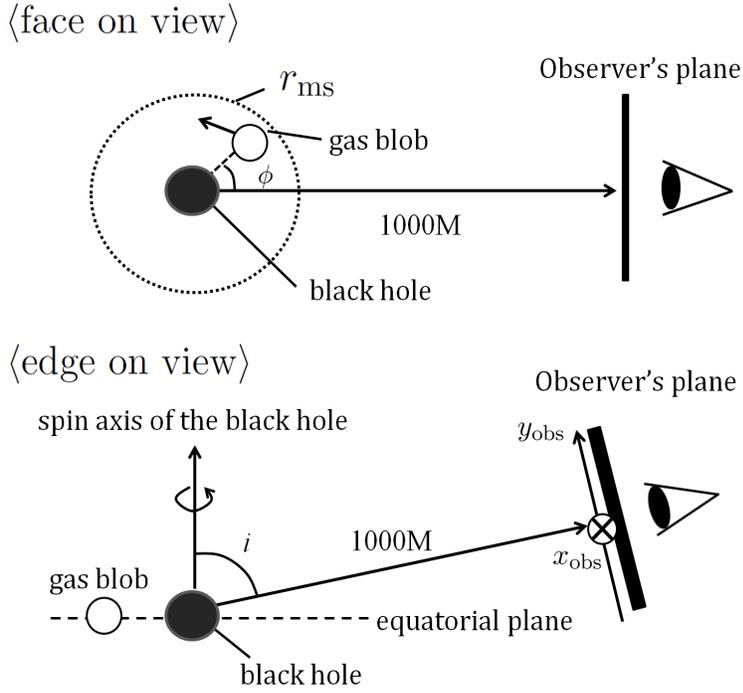} 
		\end{center}
		\caption{Schematic picture explaining the motion of the gas blob and the observer's plane.
			The $x_{\rm{obs}}$-axis is parallel to the equatorial plane of the black hole ($\otimes $), 
			and the $y_{\rm{obs}}$-axis is perpendicular to the $x_{\rm{obs}}$-axis.}
			\label{fig1}
		\end{figure}

	\subsection{Gas ring model}

	\begin{figure}
	\begin{center}
	\includegraphics[width=10cm]{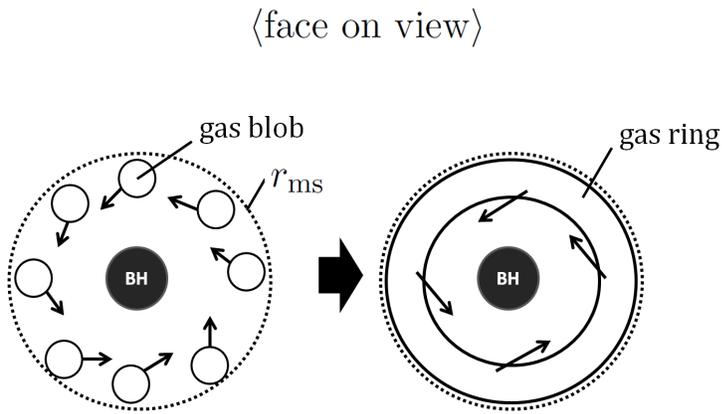} 
	\end{center}
	\caption{Schematic picture of explaining the relationship between the blob model and the ring model.}
	\label{fig2}
	\end{figure}


	Next we postulate the situation that the gas blob created in the innermost part of the disk is 
	tidally disrupted and gets elongated by the rotational velocity differences in the azimuthal direction to form a rotating ring (figure 2).
	We assume that the gas ring has the following properties: 
		\begin{enumerate}
		\item Its cross section has a circular shape and its characteristic thickness is $R_{\rm{ring}}$, which has no time dependence.
		\item We neglect self-gravity of the ring.
		\item The gas ring emits radiation with monochromatic frequency, $\nu_{0}$, in its inertial frame. 
		Its emissivity decreases with an increase of a distance from the center of the cross section of the ring 
		(hereafter referred to as the center of the ring), obeying the Gaussian function. 
		\item We neglect self absorption within the ring.
		\item The initial position of the center of the ring is at $r= 0.98r_{\rm{ms}}$.
		\item The motion of the center of the ring follows the one particle orbit given by equations (3) and (4).
		The ring falls to the black hole on the equatorial plane, keeping the constant energy, $E_{\rm ms}$, and angular momentum, $L_{\rm ms}$.
		\end{enumerate}
		
		By assumptions 1 -- 4, the emissivity, $j_{\rm{\nu}}$, in the inertial frame of the ring can be expressed as

		\noindent
		\begin{equation}
			j_{\rm{\nu}}(R)=j_{0}\exp\left[ -\left( \frac{R}{R_{\rm{ring}}} \right)^{2} \right] \delta (\nu -\nu_{0}),
		\end{equation}
		
		\noindent		
		where $R$ is a distance from the center of the ring, 
		and $R_{\rm{ring}}$, $j_{0}$, and $\nu_{0}$ are numerical constants. 
		We consider three cases for the ring thickness: $R_{\rm{ring}}=0.1M$, $0.2M$, and $0.3M$.

	\subsection{Methods of numerical calculations}	

		In order to obtain the images of infalling a blob or a ring seen by a distant observer, we solve 
		the photon trajectories by the ray-tracing method (see, e.g. Karas et al. 1992).
		The radiative transfer equation is written by

		\noindent
		\begin{equation}
			dI=g^{4}\left[\int^{\infty}_{0}j_{\rm{\nu}}d\nu\right] d\ell , 
		\end{equation}

		\noindent
		where $I$ is the intensity of a ray reaching the observer, 

		$j_{\nu}$ is the emissivity profile [equations (5) and (6)], 
		$d\ell$ is an infinitesimal spatial interval of a ray in Boyer-Lindquist coordinates, 
		and the energy-shift factor, $g$, is expressed as 

		\noindent
		\begin{equation}
		g =  \frac{1}{u^{t}(1-\Omega \Lambda )+Wu^{r}}, 
		\end{equation} 
		\noindent 
		with		

		\noindent
		\begin{equation} 
		W = -\frac{\sqrt{(r^{2}+a^{2}-a\Lambda)^{2}-\Delta [(\Lambda -a)^{2}+q]}}{\Delta}, 
		\end{equation}

		\noindent
		where $u^{\mu}$ and $\Omega(=u^{r}/u^{t})$ are 4-velocity 
		and angular velocity of the center of the blob or ring, and 
		$\Lambda\equiv L/E$ and $q \equiv Q/E$ are angular momenta 
		with respect to the $\phi$ and $\theta$ directions per unit energy, respectively.
		Note that $\Lambda$ and $q$ satisfy

		\noindent
		\begin{eqnarray}
		\Lambda &=& -x_{\rm{obs}}\sin i, \\
		q          &=& y_{\rm{obs}}^{2}-a^{2}\cos^{2}i+\lambda^{2}\cot^{2}i,
		\end{eqnarray}
		along a ray which reaches a point $(x_{\rm{obs}}, y_{\rm{obs}})$ on the observer's plane (Cunningham $\&$ Bardeen 1973).
		Here, $x_{\rm{obs}}$ and $y_{\rm{obs}}$ are 
		Cartesian coordinates on the observer's plane, 
		where the $x_{\rm{obs}}$-axis is parallel to the equatorial plane of the black hole and the $y_{\rm{obs}}$-axis is perpendicular to 
		the $x_{\rm{obs}}$-axis (see figure 1).
		Moreover, by using polar coordinates, we take 
		$(x_{\rm{obs}}, y_{\rm{obs}})=(r_{\rm{obs}}\cos\phi_{\rm{obs}},r_{\rm{obs}}\sin\phi_{\rm{obs}})$, 
		and $r_{\rm{obs}}$ and $\phi_{\rm{obs}}$ satisfy $0\leq r_{\rm{obs}}< 12M$ and $\ 0\leq \phi_{\rm{obs}}< 2\pi [\rm{rad}]$. 

		We divide the observer's plane into $n_{r}\times n_{\phi}$ cells.
		Here, the number of cells is $n_{r}\times n_{\phi}=200 \times 400 $, and spacings between each cell are 
		$dr_{\rm{obs}}=0.06M$ and 
		$r_{\rm{obs}}d\phi_{\rm{obs}}=r_{\rm{obs}}\times \pi /200$. 
		The distance between the center of the observer's plane and the black hole is set to be $r_{\rm o}=1000M$.
	
		More detailed numerical procedures are as follows:
		\begin{enumerate}
		\item We calculate the ray trajectories which leave each cell of the observer's plane 
		in the perpendicular direction to reach the vicinity of the black hole by applying the symplectic method (Yoshida 1993) obeying equations (3) and (4).
		Here the symplectic method is the numerical method to calculate test particle trajectories by solving Hamilton's canonical equation.
		We assume that a photon which enters within the radius $r=r_{\rm{h}}+10^{-4}M$ is captured by the black hole and thus cannot escape 
		from there. 
		Here, $r_{\rm{h}}$ is the radius of the event horizon of the black hole.
		
		\item Next, we calculate a trajectory of a test particle, which falls to a rotating black hole from 
		$r=0.98r_{\rm{ms}}$ by the Runge-Kutta method. This is to describe a motion of a center of the gas blob or ring. 
		We assume that a particle which enters within the radius $r=r_{\rm{h}}+10^{-4}M$ is captured by the black hole and thus cannot escape 
		from there. 
		\item We choose one ray, which reaches a certain cell of the observer's plane at an observational time, $t_{\rm{obs}}$.
		Any photons which reach the observer's plane at $t_{\rm{obs}}$ 
		after traveling time of $t_{\rm{travel}}$ along this ray 
		was radiated by the gas blob or ring at $t_{\rm{em}}=t_{\rm{obs}}-t_{\rm{travel}}$.
		At $t_{\rm{em}}$, the position and the 4-velocity of the center of the gas blob or ring are given by procedure 2.
		The emitted position of the photon which reaches the cell at 
		$t_{\rm{obs}}$ can be uniquely determined as a function of $t_{\rm{travel}}$. 
		Then by inserting the distance between the emitted position and a position of the center of the gas blob or ring into equations (5)-(6) 
		we obtain the emissivity, $j_{\nu}$, for each value of 
		$t_{\rm{travel}}$.

		\item We assume the existence of an accretion disk, 
		which is geometrically thin and optically thick, outside $r_{\rm{ms}}$. 
		That is, any photons that cross a disk surface should be absorbed by the disk 
		and do not reach a distant observer.

		\item By using $j_{\nu}$ calculated in procedure 3 and equations (7)-(11), we calculate the intensity of a ray which reaches one cell of the observer's plane at 
		$t_{\rm{obs}}$ for each value of $t_{\rm{travel}}$.
		Here, we obtain $\Lambda$ and $q$ by inserting the position of the cell into equations (10) and (11).
		
		\item We then sum up all contributions by photons which 
		were emitted at different times of $t_{\rm{obs}}-t_{\rm{travel}}$ 
		and obtain the total intensity of rays, $I$, that reaches 
		one cell at $t_{\rm{obs}}$. 
		
		\item  We finally integrate $I$ over the entire observer's plane 
		with the area, $S_{\rm{obs}}$, 
		and divide it by $S_{\rm{obs}}$ to 
		calculate flux, $f(t_{\rm{obs}})$, at each observational time, $t_{\rm{obs}}$, 
	
		\noindent
		\begin{equation}
			f(t_{\rm{obs}}) = \frac{1}{4\pi r_{\rm o}^{2}}\int I dS_{\rm{obs}},
		\end{equation}
		where $dS_{\rm{obs}}$ is the area of the cell. 
		We define $t_{\rm{max}}$ as the time when $f(t_{\rm obs})$ reaches its maximum, 
		and then, the normalized flux, $F(t_{\rm obs})$, is 
	
		\noindent
		\begin{equation}
		F(t_{\rm{obs}}) = \frac{f(t_{\rm{obs}})}{f(t_{\rm{max}})}.
		\end{equation}
		Furthermore, by using $j_{\rm{\nu}}d\ell$ at a radiative position, we calculate a photon number $N(t,g)\delta g$ 
		whose energy-shift factor is within $g \sim g+\delta g$ at time, $t_{\rm{obs}}$: 
		
		\noindent	
		\begin{equation}
		N(t_{\rm{obs}},g)\delta g =\frac{1}{h\nu_{0}}\int \int g^{3}j_{\rm{\nu}}d\ell dS_{\rm{obs}} \delta g,
		\end{equation}
	
		\noindent	
		where $h$ is Planck's constant, and the integration is made to satisfy that the energy-shift factors of 
		the rays reaching the observer are within a range between $g \sim g+\delta g$.
		\end{enumerate}
		
		Hereafter, we re-write the observational time, $t_{\rm obs}$, simply as $t$.
	
\section{Results}

	\subsection{Blob model: typical case}

			\begin{figure}
				\begin{center}
					\includegraphics[width=14cm]{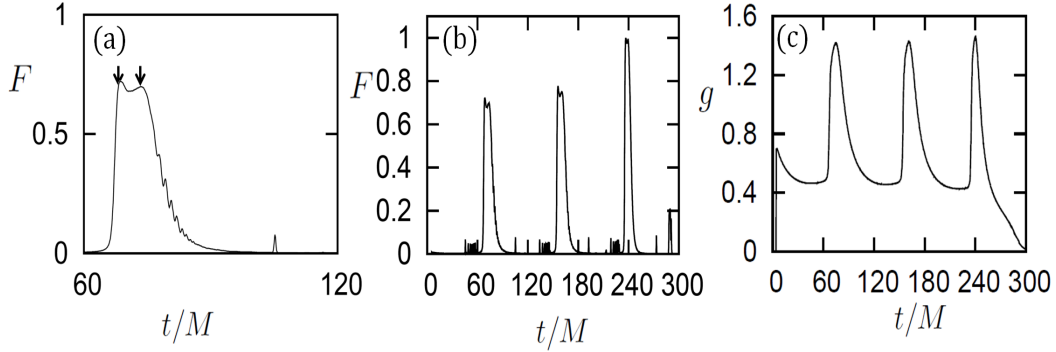}
				\end{center}
					\caption{Typical results of the blob model for the case with $(a/M,i)=(0.0, 85^{\circ})$ and $R_{\rm{blob}}=0.1M$: 
								(a) the detailed light curve around the first peak, 
								(b) the entire light curve, and 
								(c) the time variation of the centroid energy-shift, 
								where contributions by the photons which reach the observer 
								after crossing the equatorial plane 
								are removed in this plot.
								}\label{fig3}

			\end{figure}

				We first show in figure 3 the results of a typical case 
				in which we assign $(a/M,i)=(0.0, 85^{\circ})$ and $R_{\rm{blob}}=0.1M$.
				The detailed light curve around the first peak shown in panel (a) exhibits 
				two peaks [see arrows in panel (a)]. 
				The former peak appears, when the blob is at the opposite side to 
				the observer's position (phase $\phi=\pi$, see figure 1), 
				due to the gravitational lensing [effect (I) in table 1].
				The latter peak is due to the beaming occurring at $\phi=1.5\pi$ 
				[effect (II) in table 1].
				After the second peak, the flux dumps by the redshift due to 
				the motion of the blob moving away from the observer.
				At the time of the second peak, a centroid energy-shift factor reaches 
				its maximum [see panel (c) of figure 3]. 
				These properties are the same as the those studied of a rotating blob 
				(cf. Karas et al. 1992; $\rm{Dov\check{c}iak} $ et al. 2004a). 

				One might say, therefore, that the light variation in the first stage of evolution looks 
				quite similar to that of a rotating blob at a fixed radius. 
				Note, however, one important difference between them; 
				the peak intensity gradually increases with time in our model [see figure 3(b)]. 
				Such an increase occurs, when a blob approaches the radius of 
				the photon circular orbit, $r_{\rm{ph}}$, and this phenomenon is caused by 
				the focusing effect around the photon circular orbit [effect (III) in table 1].

				Let us explain its meaning of the focusing effect around $r_{\rm ph}$ in a simple case. 
				By inserting $\theta=90^{\circ}$ and $a/M=0$ in equations (3) and (4) we find the $r$ component of the geodesic equation to be 
				[see also Shapiro $\&$ Teukolsky (1983), section 12.7]: 
				\begin{eqnarray}
				\left(\frac{dr}{d\lambda}\right)^{2}=\frac{1}{\Lambda^{2}}-\frac{1}{r^{2}}\left( 1-\frac{2M}{r}\right), \nonumber
				\end{eqnarray}
				where $\Lambda$ is given by equation (10). 
				It is easy to show that the second term on the right-hand-side of this equation reaches its maximum at $r_{\rm ph} = 3M$. 
				That is, when a ray travels from inside $r_{\rm ph}$, the value of $dr/d\lambda$ first decreases until $r_{\rm ph}$ 
				where it reaches its minimum, and then increases outward. It then follows that the rays emitted isotropically around 
				$r_{\rm ph}$ are bound around the circular orbit for a while, as is illustrated in figure 4, 
				and then, reach the observer at nearly the same time. 
				This is why the observed luminosity increase occurs in figure 3(b). 
 				
				\begin{figure}
					\begin{center}
						\includegraphics[width=9cm]{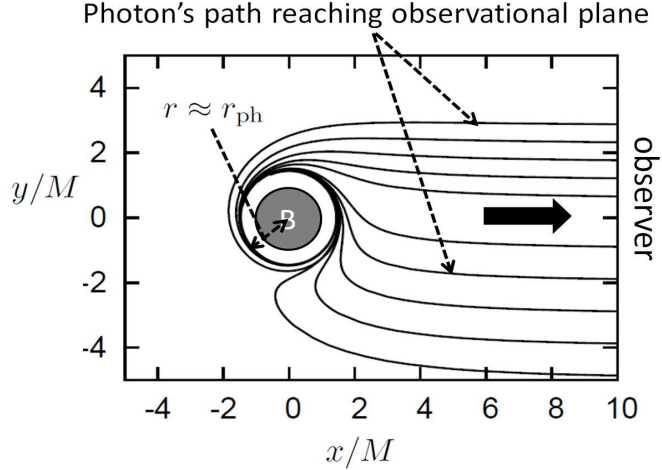}
					\end{center}
						\caption{Schematic picture explaining the focusing around the photon circular orbit, $r_{\rm{ph}}$ ($a/M=0.9$).
						Each line represents a ray that finally reaches the observer's plane. 
						}\label{fig4}
				\end{figure}


				Figures 3(b) and (c) clearly prove the existence of the focusing effect around $r_{\rm ph}$. 
				The flux reaches its maximum at $t_{\rm{max}}$, and finally decreases with time, 
				
				while the centroid energy-shift factor of the peak is nearly constant. 
				That is, the peak flux increases in the first stage 
				because of an increase of photon numbers, not because of 
				an increase of mean photon energy; 
				which provides a support to the idea that flux increase is due to 
				the focusing of rays.
				In the second stage (after $t_{\rm max}$), the flux decays due to the gravitational redshift [effect (V) in table 1] 
				and to an increase in the fraction of photons that are captured  by the black hole.

			\subsection{Blob model: cases with various $(a,i)$}
				
				Next, the results of other cases with various combinations of 
				spin parameters  and inclination angles are shown in figure 5.

				We first fix $i$, say, $i=85^\circ$, and let us compare the non-spin case
				[$a/M=0.0$; figure 5(a1)], middle one [$a/M=0.6$, figure 5(a2)] 
				and high spin one [$a/M=0.9$, figure 5(a3)]. We immediately notice that
				the higher the spin is, the shorter becomes the peak interval.
				This is because as $a$ increases, $r_{\rm ms}$ decreases,
				and, hence, the angular frequency ($\Omega$) increases, 
				whereas the orbital period ($\propto \Omega^{-1}$) decreases. 
				One may think that we can easily estimate $a$ by measuring peak intervals
				on the condition that we precisely know $M$.
				This is, however, not so feasible in practice, since
				it is hard to detect radiation from one separate, compact blob.
				Rather, it is more likely that assembly of blobs (or a ring) fall together. 
				Therefore, we had better pay more attention to the overall shape of the light curves.
				If we see how the maximum flux of each peak changes with time in figure 5,
				we find that the higher the spin is, the more becomes the number of peaks,
				and the more rapidly grows the maximum flux of the peak.
				We also notice a similar tendency in the time variations of $g$
				displayed in the lower panels. These facts lead us to the conclusion
				that how the peak flux and the energy-shift factor vary with time
				can be good indicators of the black hole spin.

				Next, we fix $a$ and change $i$, finding that
				the flux peak increase is more appreciable in high $i$ systems.
				In fact, the first stage evolution (when the peak flux increases) is
				clear when $i=85^\circ$, while it is not when $i=20^\circ$. 
				This is because the focusing effect around $r_{\rm ph}$ 
				is more appreciable in nearly edge-on systems
				with smaller $i$ values.
				Further, the higher $i$ is, the sharper becomes each peak in the $g$ variation
				profile (see lower panels).
				This can be naturally understood, since the beaming effect is more effective
				in higher $i$ systems.

				\begin{figure}
					\begin{center}				
						\includegraphics[width=14cm]{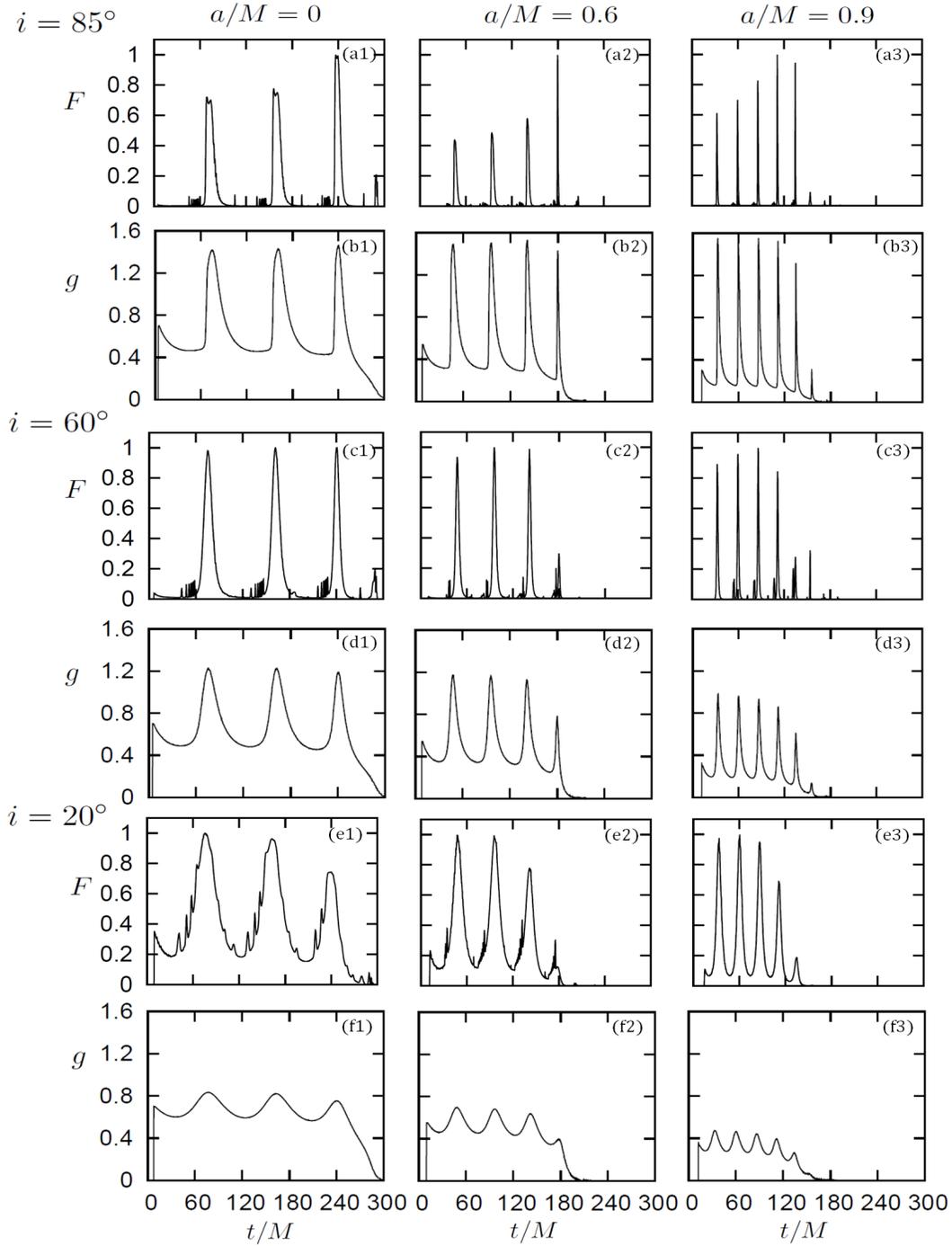}
					\end{center}						
						\caption{Time variations of the normalized flux, $F(t)$, (upper panels) 
									and those of the centroid energy-shift factors, $g(t)$, 
									(lower panels) for various combinations of $a$'s and $i$'s.
									The abscissa is observational time $t$ normalized by the black hole mass $M$.
									}\label{fig5}
				\end{figure}
				\clearpage

In order to extract information regarding the spin ($a/M$) from the overall 
flux variations shown in figure 5, we next suppose the cases 
in which numerous blobs fall together.
We plot in figure 6 superposition of the light curves of numerous blobs 
rotating with different initial orbital phases for the case of $a/M=0.0$ (panel a1), 
0.6 (a2), and 0.9 (a3), respectively.
Here, we fix the inclination angle to be $i=85^\circ$.
It is evident that the higher $a$ is, the shorter becomes the time to reach 
the flux maximum. This is because the initial position of the blob 
(at $\approx r_{\rm{ms}}$) is closer to the black hole with a higher $a$ value. 

In every case, two stage evolution (rise and decay) is clear.
Such a trend can be more clearly shown in panels b1 -- b3 in figure 6,
in which envelope curves connecting each peak in panels a1 -- a3 are displayed.
To be more precise, the first stage in figure 6(b1) can be further divided to 
two sub-stages; that is, the flux first slowly increases (until $t/M\sim 220$)
and then starts to rapidly increase until the flux maximum at $t/M\sim 270$,
followed by a rapid decay at the end.
Such sub-stages are not so clear in high spin cases [see figure 6(b3)].

Further, we wish to draw attention to 
the $a$-dependence of the shape of the flux curve around its maximum;
the peak is rather broad when $a/M$ is high [see panel (b3)],
while it is sharp when $a/M$ is low [see panel (b1)].
That is, the region where focusing is prominent widens as $a/M$ increases.
This can be understood by the consequence of the frame dragging around 
a rotating black hole, by which photons are focused in the rotating direction [effect (IV) in table 1].

In the high $a$ case, the frame dragging effect plays an important role for flux peaks in panels (b). 
The each radius at which point the each flux peak is $r/M=(3, 2.9, 2.1)=(1,1.32,1.34) r_{\rm ph}$, for $a/M=(0,0.6,0.9)$, respectively. 
In the case of $a/M =0$, the emitting radius for the flux peak is equal to $r_{\rm ph}$. 
In the case of $a/M = 0.6$ and $0.9$, on the other hand, the radii of the flux peaks are outside of $r_{\rm ph}$. 
This can be understood by the consequence of the frame dragging around a rotating black hole, by which photons are focused in the rotating direction.

We should make cautions here, however,
since what are observed are not the envelope curve of the peaks 
but the total radiation energy received by an observer per unit time.
We thus add three more panels as figures 6(c1) -- (c3) 
displaying the time evolution of ``peak area'',
time integral of radiation energy of the peak. 
By comparing the bottom panels (c1) -- (c3) with the middle panels 
(b1) -- (b3), we understand that the sharp rise found in the
middle panels are totally missing in the bottom panels and that
such a distinction is clearer in lower spin cases.
Since the bottom panels show the time integration of flux of each peak, 
while the middle panels show its maximum flux only, differences between 
them should mean differences in time width (i.e., duration) of each peak. 
We actually find that the time width of the peak significantly decreases 
with time because of the Doppler beaming effect is more enhanced 
when gas moves nearer to the black hole. 
The bottom panels will be used for our method of spin measurements.

			\begin{figure}
				\begin{center}
					\includegraphics[clip,width=12.5cm]{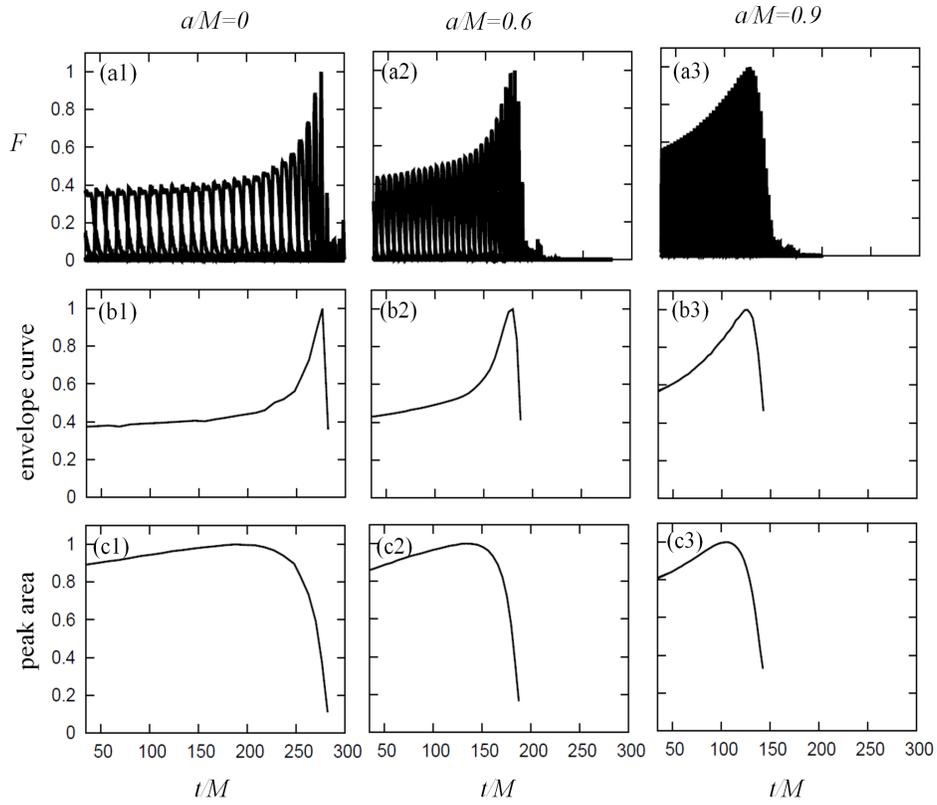}						
				\end{center}
					\caption{Flux variations of numerous blobs for cases with different spin parameters,
								$a/M=0.0$, $0.6$ and $0.9$.
								Panels (a) show superposition of light curves of numerous blobs
								rotating with different initial orbital phases.
								Panels (b) plot envelope curves connecting the peaks shown in panels (a).
								Panels (c) depict time developments of the peak areas, time integral of
								the peaks shown in panels (a). Here we fix $i=85^{\circ}$.
								The abscissa is observational time $t$ normalized by the black hole mass $M$, 
								and the ordinate is the flux normalized by its maximum value in each panel.
								}\label{fig6}
			\end{figure}
			\clearpage

	\subsection{Results of ring model: light curves}

			\begin{figure}
				\begin{center}
					\includegraphics[width=14cm]{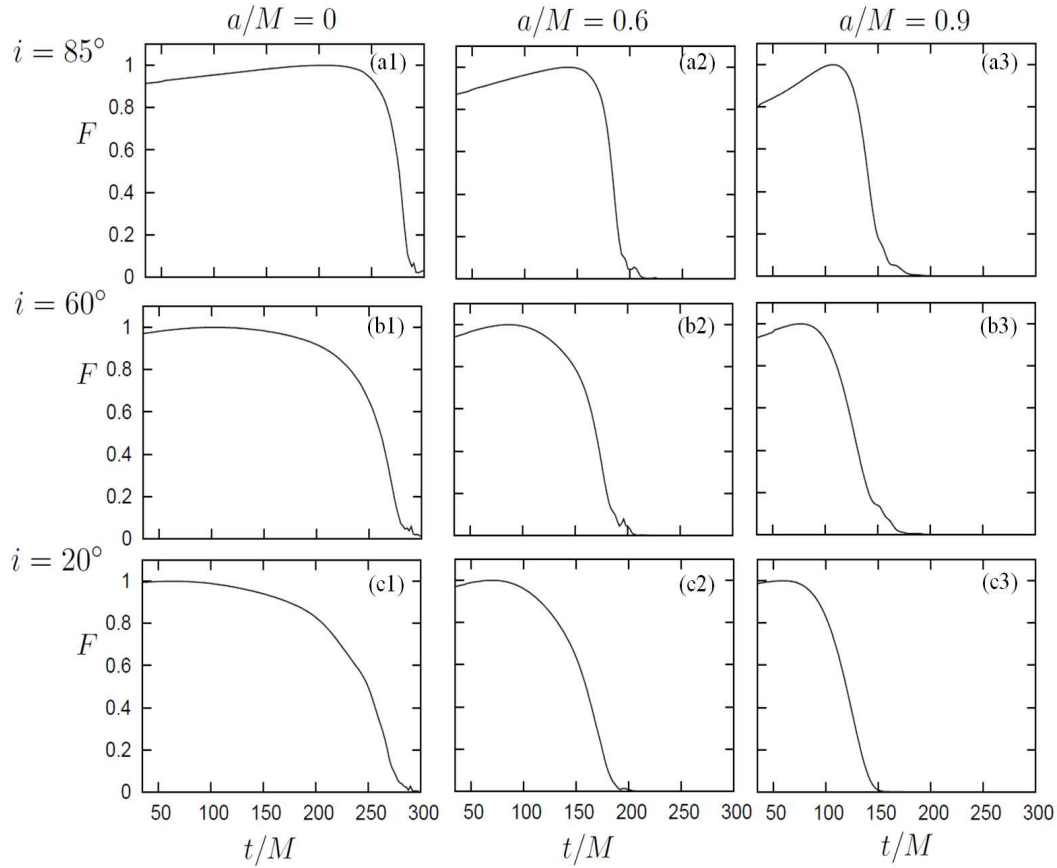}						
				\end{center}
					\caption{Light curves of the ring for the spins of $a/M=0.0$ (left), 
								$0.6$ (middle), and $0.9$ (right) and the inclination angles of 
								$i=85^{\circ}$ (upper), $60^{\circ}$ (middle), and $20^{\circ}$ (lower), 
								respectively. The abscissa is observational time normalized by $M$, 
								and the ordinate is the flux normalized by its maximum value, $F(t)$.
								We fix $R_{\rm{ring}}=0.1M$.
								}\label{fig7}
			\end{figure}

				We show the light curves of the ring for various combinations of $a$ and $i$ 
				for a fixed value of $R_{\rm{ring}}=0.1M$ in figure 7. 
				The continuum curves are nothing but the superpositions of the light curves of 
				numerous blobs with different initial orbital phases [cf. panels (c) of figure 6].

				There are several noteworthy features found in these light curves. 
				First, the light curves have no peaks unlike the case of the blob model. 
				This is because a distant observer at any time receives the ray
				which is at the orbital phase of $\phi \sim \pi$ (at which 
				the gravitational lensing is most effective) or the one at $\phi \sim 1.5\pi$ 
				(at which the beaming is most effective).
				Next, the flux first increases due to the focusing effect around $r_{\rm ph}$
				(in the first stage) 
				and then decays due to the gravitational redshift and to the capture of 
				rays by the black hole (in the second stage).
				These features are the same as those of the numerous blobs shown in figure 6.

				Let us examine the $(a,i)$-dependence of the light curves. 
				Let us first fix the inclination angle ($i$) and change the spin 
				from $a/M=0.0$ to $a/M=0.9$ to see what changes do occur.
				\begin{enumerate}
				\item The higher $a$ is, the shorter is the flux variation 
				for all the values of $i$.
				This is because a ring falls to the black hole from smaller radii 
				(recall that the initial position at $r \approx r_{\rm{ms}} $ is smaller
				for a higher spin).
				\item The higher $a$ is, the steeper is the flux curve in the first stage. 
				This feature is prominent for high $i$ cases; e.g., 
				$i=85^{\circ}$ and $60^{\circ}$, 
				Note that flux increase itself is not so appreciable when $i$ is low. 
				This is because the focusing effect around $r_{\rm ph}$, which is responsible 
				for the flux increase, is stronger when $a$ is high and $i$ is higher. 
				\end{enumerate}

				Next, we fix $a/M$ and change the inclination angle from
				$i=20^{\circ}$ to $i=85^{\circ}$.
				As stated above, the flux increases more rapidly, in higher $i$ cases
				because of more enhanced the focusing effect around $r_{\rm ph}$.
				Same feature is seen in other spin cases.

			\subsection{Results of ring model: photon numbers} 
			
				\begin{figure}
					\begin{center}
						\includegraphics[width=14cm]{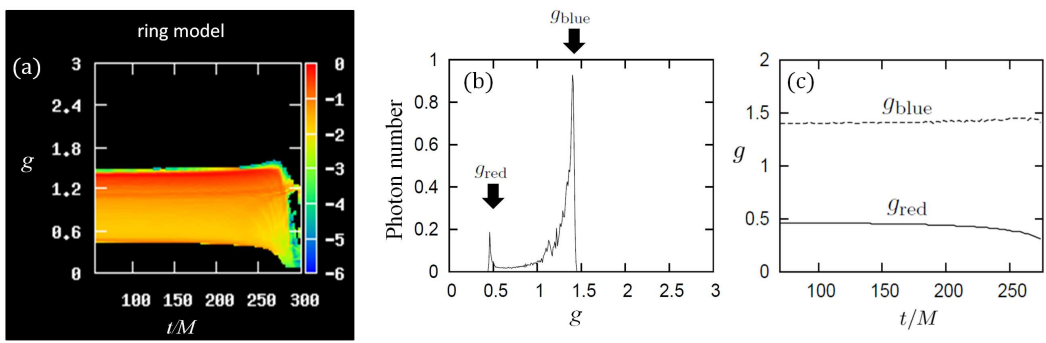}		
					\end{center}
						\caption{
						Time-dependent properties of the radiation emitted from a ring.
						We fix $(a/M,i)=(0.0,85^{\circ})$ and $R_{\rm ring} = 0.1M$. 
						Panel (a) shows the time variations of the photon numbers as a function of the energy-shift factor, $g$.
						The color represents a logarithm of the photon numbers normalized by their maximum value.
						Panel (b) shows the photon number distribution against $g$ at the fixed time of $t/M=50$. 
						Two peaks are prominent, which are indicated by the two arrows. 
						We call $g_{\rm{blue}}$ (or $g_{\rm{red}}$) as the $g$-value of the blue-shifted (red-shifted) component. 
						Panel (c) shows the time variations of $g_{\rm{blue}}$ and $g_{\rm{red}}$. 
						}\label{fig8}
				\end{figure}

				Although the flux variation contains useful information regarding the spin and 
				the inclination angle, it is not sufficient to measure the both quantities.
				Another distinct information is needed. 
				Since two pieces of independent information are mixed in radiation flux, $F(t)$:
				photon energy and photon numbers, it may be useful to separate them.
				We next investigate the $(a,i)$-dependence of the energy-shift factor, $g$.

				Time variation of the photon number is displayed in figure 8(a).
				Here, the color represents a logarithm of the photon number normalized by 
				its maximum value. 
				We here fix $(a/M,i)=(0.0,85^{\circ})$ and $R_{\rm ring} = 0.1M$.
				Panel (b) shows the photon number profile (against $g$) at a fixed time of $t/M=50$. 
				Two peaks are prominent: one is blue-shifted peak at $g\sim 1.4$ 
				and another is red-shifted peak at $g\sim 0.5$ (see arrows). 
				Hereafter, we call $g_{\rm{blue}}$ (or $g_{\rm{red}}$) as the $g$-value of the blue-shifted (red-shifted) component.

				To consider the reason for the two peaks, 
				let us see again the time variation of the centroid energy-shift  of an infalling blob shown in figure 3(c).
				This panels exhibit two key features (or phases): 
				(i) sharp peaks with large $g$-values (greater than $1.0$), 
				which occur due to the beaming, and 
				(ii) long {\lq}quiescence' with low $g$-values (typically $g\sim 0.5$),
				which occurs when the blob is moving away from the observer
				due to the Doppler effect and to the gravitational redshift.
				Now we understand that
				$g_{\rm blue}$ (or $g_{\rm red}$) in figure 8(b) corresponds to the centroid energy-shift factor 
				of the peaks (or of the quiescence) in figure 3(c). 
				Here, we wish to stress that figure 8(b) shows a local maximum at $g\sim g_{\rm{red}}$, although the photon energy is low. 
				This means that the area under the F(t) curve of the red-shifted 
				component is larger than those of other energy-shift factors.
 
				The double-horn profile displayed in figure 8(b) resembles that of the `disk line' 
				Fe K$\alpha$ emission line formed on an relativistic accretion disk around a black hole 
				(Kojima 1991; Laor 1991). We wish to point a big distinction, however, that 
				our red peak is very narrow, while the red component of the disk line is usually broad, 
				especially when $a/M$ is large. This is because we are concerned with emission
				from a narrow ring ($R_{\rm ring} \ll r$), while the disk line is produced on a disk 
				with a large area. Superposition of emission lines from different radii produces 
				the broad red wing of the disk line.

				Finally the time variations of these quantities are shown in Panel (c).
				The spin and inclination-angle dependence is expected, which will be shown in 
				next subsection. 

			\subsection{Results of ring model: energy-shift factor}

				\begin{figure}
					\begin{center}
						\includegraphics[clip,width=14cm]{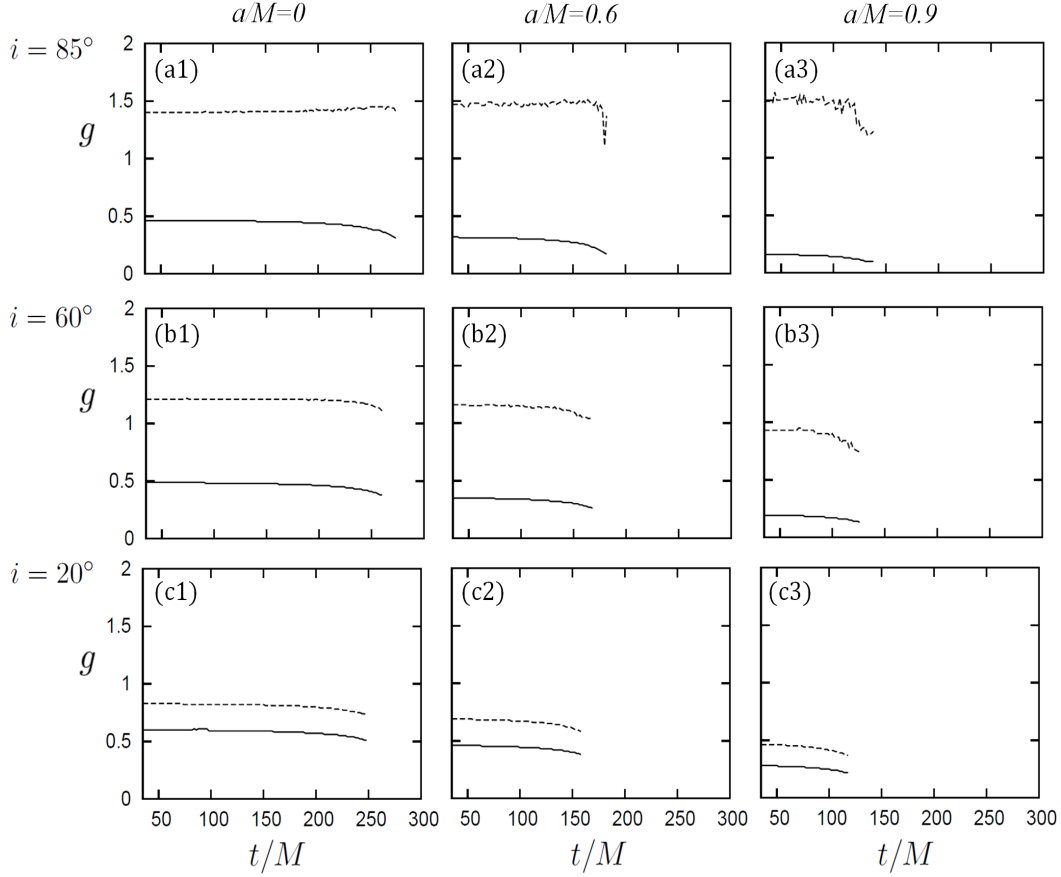}		
					\end{center}
						\caption{Time variations of $g_{\rm{blue}}$ (dot line) and $g_{\rm{red}}$ (solid line)
									for a variety of $a/M=0.0$ (left), $0.6$ (middle), and $0.9$ (right) and
									$i=85^{\circ}$ (upper) ,$60^{\circ}$ (middle), and $20^{\circ}$ (bottom), respectively.
									Each curve is plotted until the time when the normalized flux decays to reach $F=1/2$ 
									(i.e., half of the maximum flux).
									}\label{fig9}
				\end{figure}
				
				We show the time variations of $g_{\rm{blue}}$ and $g_{\rm{red}}$ for various combinations of 
				$a$ and $i$ in figure 9.
				First, we immediately notice a clear trend
				that $g$-values depend more sensitively on $i$, rather than $a$, and that 
				the higher $i$ is, the longer becomes the energy-shift interval between the two lines. 
				This can be easily understood, since the $Doppler$ is more pronounced when 
				$i$ is higher. 
				Let us next fix $i$ and see how the results change by decreasing $a$ from 
				high $a/M (=0.9)$ values to lower one. Here is a summary:
				(i) When $i$ is low; e.g., $i=20^{\circ}$, 
				both of $g_{\rm{blue}}$ and $g_{\rm{red}}$ decrease with $a$ at low $i$,
				since gravitational redshift is stronger in order that the initial position of the ring 
				$(\approx r_{\rm{ms}})$ is smaller [see figures 9(c1) and (c3)].
				(ii) When $i$ is large; say, e.g., $i=85^{\circ}$, 
				$g_{\rm{blue}}$ does not depend so much on $a$,
				while since the beaming effect is stronger than the gravitational redshift,
				while $g_{\rm{red}}$ is smaller due to the gravitational redshift and the redshift 
				due to the motion of the ring fraction moving away from the observer.

\section{New method of black-hole spin measurement}			
\subsection{Introduction of new quantities}

As is demonstrated in the previous section, radiation from an infalling gas ring
contains useful information for the $a$ and $i$ measurements of a central black hole.
In this section we propose a methodology for the measurement of $a$ and $i$.
In the proposed method we use the following three quantities:
(i) $g_{\rm{red}}(t)$, energy-shift factor of the red-shifted spectral component, 
(ii) $g_{\rm{blue}}(t)$, the same but of the blue-shifted component, and 
(iii) $F(t)$, the normalized radiation flux (see figures 7 and 9).
From these quantities we calculate the following six quantities:
\noindent			
\begin{eqnarray}
A_{1} &=& \frac{2}{M}\int^{t_{\rm{max}}}_{t_{\rm{i}}}\frac{F(t)-F(t_{\rm{i}})}{1-F(t_{\rm{i}})}dt,\\
				A_{2} &=& \frac{4}{3M}\int^{t_{\rm{f}}}_{t_{\rm{max}}}F(t)dt,\\
				A  &=& A_{1}+A_{2},  \\
				B &=& \frac{\int^{t_{\rm{f}}}_{t_{\rm{i}}}g_{\rm{red }}(t)dt}{(t_{\rm{f}}-t_{\rm{i}})} = \langle g_{\rm{red}}\rangle,\\
				C &=& \frac{\int^{t_{\rm{f}}}_{t_{\rm{i}}}g_{\rm{blue}}(t)dt}{(t_{\rm{f}}-t_{\rm{i}})} = \langle g_{\rm{blue}}\rangle,\\
				D &=& \frac{C}{B}-1, 
\end{eqnarray}
\noindent 
where we fix the initial time to be $t_{\rm{i}}/M=60$ 
and $t_{\rm{f}}$ is defined in such a way that $F(t_{\rm{f}}) \cong 0.5$. 
Note also that $t$ is normalized by $M$. 
One may think that this definition of $t_{\rm i}$ looks quite arbitrary
and that the value of $A_1$ crucially depends on how to choose $t_{\rm i}$.
Fortunately, however, this is not the case,
since only the flux which is substantially greater than $F(t_{\rm i})$ 
contributes to the integral in $A_1$. 
In other words, $A_1$ depends not critically on $t_{\rm i}$ but on $t^\prime_{\rm i}$, 
when the flux starts to grow rapidly.

Let us next consider the physical meaning of $A_1$, $A_2$ and $A$
by taking an example of a simple functional form for $F(t)$.
Namely, we assume that $F(t)$ stays nearly constant $F(t)\sim F(t_{\rm i})$ until
$t^{\prime}_{\rm i}$ and that it grows as a linear function of $t$ until $t_{\rm max}$;
\begin{equation}
  F(t) \approx F(t_{\rm i}) + [1 - F(t_{\rm i})] \frac{t-t^\prime_{\rm i}}{t_{\rm rise}},
\end{equation}
where $t_{\rm rise}\equiv t_{\rm max} - t^\prime_{\rm i}$ is the typical rise time. 
We then find
\begin{equation}
  A_1 \approx t_{\rm rise}/M=(t_{\rm max} - t^\prime_{\rm i})/M.
\end{equation}
Note that $A_1$ does not depend on the initial flux, $F(t_{\rm i})$ nor $t_{\rm i}$.
If we assume the linear functional form for $F(t)$ also for a decay,
\begin{equation}
  F(t) \approx F(t_{\rm f}) + [1-F(t_{\rm f})] \frac{t_{\rm f}-t}{t_{\rm decay}},
\end{equation}
where $t_{\rm decay}=t_{\rm f} - t_{\rm max}$ is the typical decay time, we obtain
\begin{equation}
  A_2 \approx (t_{\rm f} - t_{\rm max})/M
\end{equation} 
We finally have 
\begin{equation}
  A = A_1 + A_2 \approx (t_{\rm f} -t^\prime_{\rm i})/M
\end{equation} 
Thus the value of $A$ means the typical timescale of flux variation. 

In table 2 we summarize the physical meanings of the quantities, $A$-$D$. \\
\begin{table}
  \tbl{Physical meanings of the quantities, $A$--$D$.}{%
  \scalebox{1.2}[1.2]{
	\begin{tabular}{clc}\hline quantity & physical meaning	& equation number \\ \hline
	$A$	& timescale of flux variation		  & (17) and (25)   	\\ 
	$B$	& time averaged $g_{\rm{red}}(t)$         & (18)		\\ 
	$C$	& time averaged $g_{\rm{blue}}(t)$ 	  & (19)		\\ 
	$D$	& relative difference between $B$ and $C$ & (20)    	  	\\ \hline
    \end{tabular}}}\label{tab:2}
\begin{tabnote}
\end{tabnote}
\end{table}

In the next section, 
we calculate $A_{1}$--$D$ for a variety of combinations of $a$, $i$, and $R_{\rm{ring}}$
to seek for the best combinations of the quantities to extract information 
on the spin from observational data.

\begin{figure}
\begin{center}
\includegraphics[clip,width=10cm]{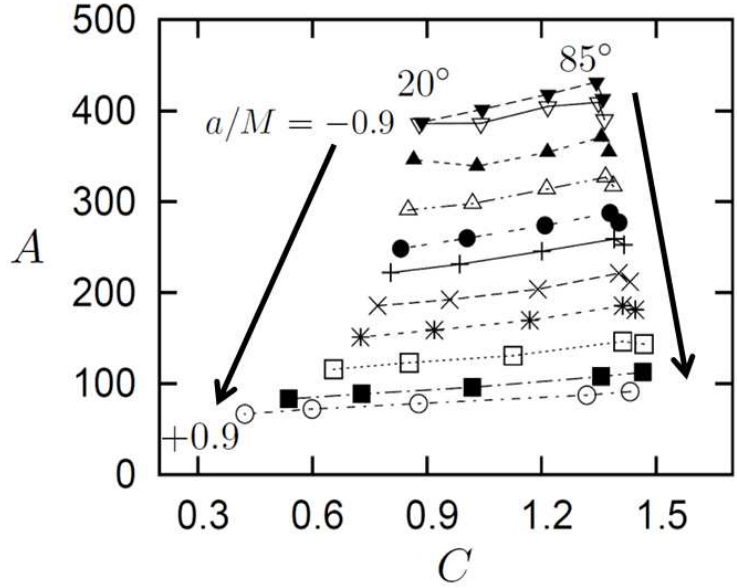}
\end{center}
\caption{Relation between $A$ and $C$ 
for the inclination angles of $i$=$20, 40, 60, 80$ and $85^{\circ}$
and for the spin parameters of $a/M=-0.9, -0.8, -0.6, -0.4, -0.2, 0, 0.2, 0.4, 0.6, 0.8$ and $0.9$.  
We fixed the ring width to be $R_{\rm{ring}}=0.1M$.
The data points with the same $a$ value are connected by the lines. 
The two arrows indicate the direction of increasing $a$.
}\label{10}
\end{figure}

\begin{figure}
\begin{center}
\includegraphics[clip,width=14cm]{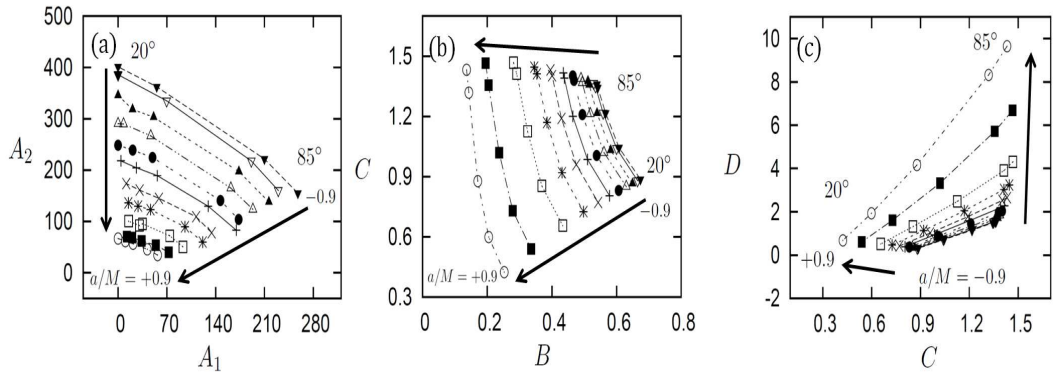}
\end{center}
\caption{Relation between the quantities except for $A$ and $C$  
for the same inclination angles, spin and ring width on figure 10.
}\label{11}
\end{figure}
			
\subsection{Dependence of $A_{1}$--$D$ on $(a,i)$}	

The calculation results are summarized in figure 10 and 11. 
It is important to note that all the curves in each panel do not cross each other, that is; 
we can, in principle, uniquely determine $a$ and $i$, once the values of $A_{1}$--$D$ are specified.

Particularly, we notice that the lines (on which $a$ is constant) are nearly horizontal;
that is  $A$ does not so much depend on $i$ (figure 10). 
This fact leads to an important conclusion that we can roughly estimate $a$ solely by $A$.
We emphasize that it shows a way to break the classic $a$-$i$ degeneracies of the spectral methods of
measuring spin, and is not particularly model dependent like most QPO methods.

Let $a/M$ increase from $0.0$ to $0.9$ for a fixed value of $i$ to see
how the key quantities of $A_{1}$--$D$ vary.
\begin{enumerate}
\item 
We find that all of the $A$ values decrease with an increase of $a/M$.
This is because the timescale of flux variation decreases with an increase of $a/M$ 
both in the rise and decay phase for all $i$ (see figure 7). 

\item With an increase of $a/M$ both of $B (=\langle g_{\rm{red}}\rangle)$
and $C (=\langle g_{\rm blue}\rangle)$ decrease, 
unless $i$ is very high, because of the gravitational redshift (see figure 9).
Since the decrease of $C$ is slower than that of $B$, 
$D[\equiv (C/B)-1]$ increases with an increase of $a/M$. 
When $i$ is very high (e.g., $i=85^\circ$), conversely, 
$C$ rather increases (while $B$ still decreases) with an increase of $a/M$,
This is due to the beaming effect, which is more effective for edge-on cases
\end{enumerate}

\subsection{Dependence of $A_1$--$D$ on the thickness of the ring}

In the previous subsection, we propose a new method for the spin measurement.
One may think, however, that the results may sensitively depend on the ring shape, 
such as the characteristic thickness of the ring. To demonstrate that this is not the case,
we need to consider how the values of $A_{1}$--$D$ depend on the thickness of the ring.
We will demonstrate that the results will not depend on $R_{\rm ring}$
as long as if we use $A$, instead of $A_1$ or $A_2$.

We display in figure 12 the relation between several quantities for
$R_{\rm{ring}}=0.2M$ [panels (a1) and (a2)] and $0.3M$ [panels (b1) and (b2)], respectively. 
By comparing these panels with figures 11(a) and 10, we understand that 
the relation between $A$ and $C$ do not critically depend on the thickness of the ring.
Since we put the center of the ring (where the emissivity $j_{\nu}$ is maximum)
at the same position for all the models, we can easily understand 
that $B$--$D$ do not so much depend on a thickness of the ring.
For low inclination angles (say, $i=20^{\circ}$), hence, 
 we can determine $a$ and $i$ by using $B,C$ and $D$. 
For high inclination angles (say, $i >60^{\circ}$), however, 
the photon number of the red-shifted component is much less than that of the blue-shifted one
so that the precise measurements of $g_{\rm{red}}$ from actual observations might be difficult.
From henceforth we mainly use the information $A$ and $C$.

Figures 12(a1) and (b1) display how the relation between $A_{1}$ and $A_{2}$ depend on
the thickness of the ring changes. Here, we fixed $a$ and $i$.
Comparing these with figure 9(a) showing the case with $R_{\rm{ring}}=0.1M$, we see
a clear tendency that
the larger $R_{\rm{ring}}$ is, the larger is $A_{1}$ and the smaller is $A_{2}$.
This can be understood, since a larger thickness means a longer emission timescale,
thereby $A_1$ being increased.
(Note that the motion of the center of the ring is kept the same regardless of the ring thickness, 
but the epoch when the last tip of the ring material passes a certain radius (say, $r_{\rm ph}$) is
more and more delayed, as the ring thickness increases.)
Since the decay timescale decreases, so does $A_2$, as $R_{\rm ring}$ increases.

To summarize, the thicker the ring is, the larger $A_{1}$ is and the smaller $A_{2}$ is,
whereas $A$ ($=A_1+A_2$) does not critically depend on the ring thickness 
[see figure 10, figures 12(b1) and (b2)]. 
In fact, the relation between $A$ and $C$ shown in figures 12(a1), (b1)
and figure 11(a) exhibit similar tendencies among different ring models.

Let us finally demonstrate that we can certainly have good guesses for $a/M$ and $i$, 
by using the $(A, C)$ diagram in figure 12. 
The results are summarized in table 3. 
We find that the differences in $a/M$ and $i$ among different models are within ranges of $\sim 0.01-0.1$ 
and $\sim 1^{\circ}$, respectively.
To conclude, we can measure the spin and inclination angle with good accuracy by using $A$ and $C$. 

\begin{figure}
\begin{center}
\includegraphics[width=14cm]{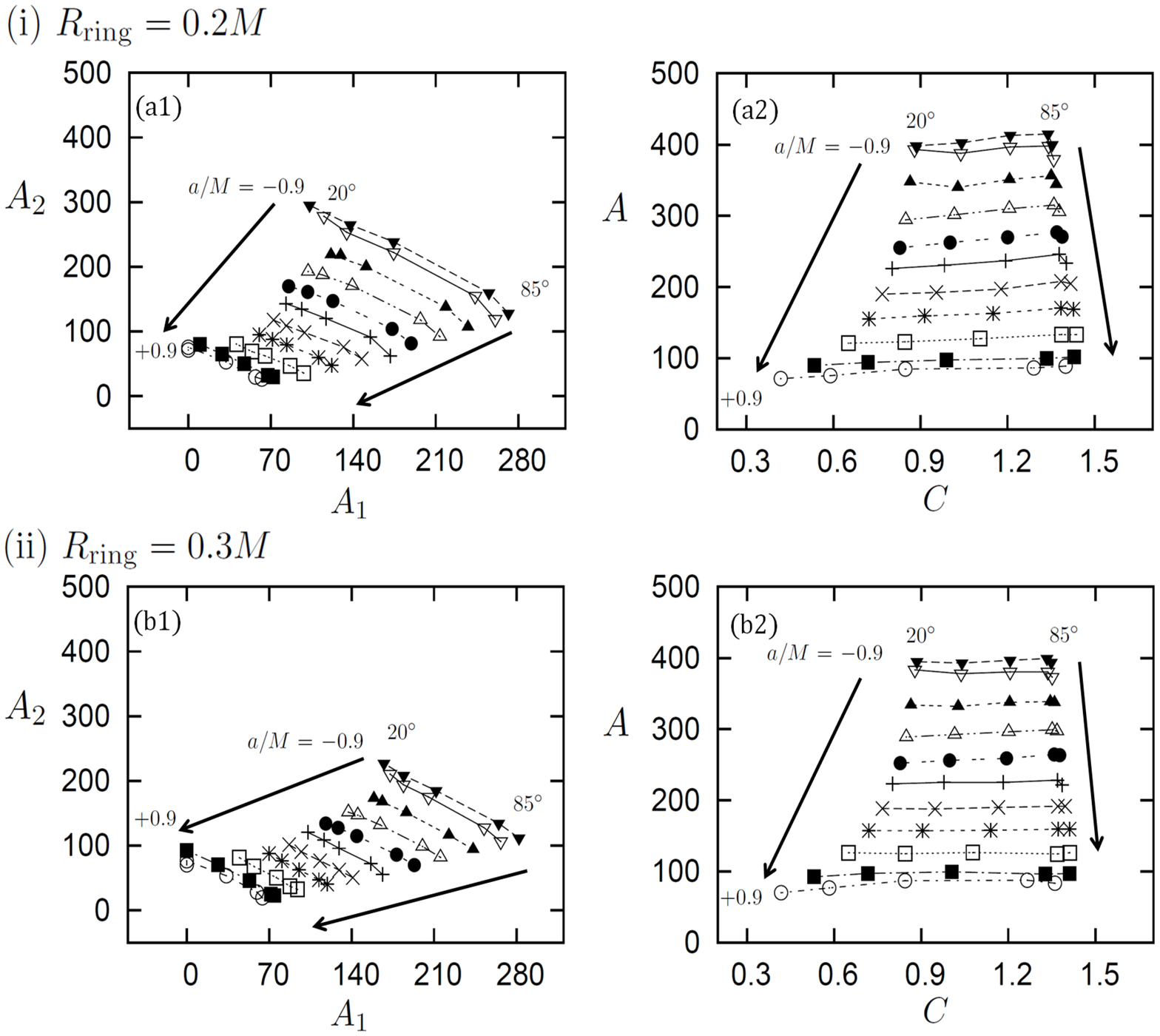}					
\end{center}
\caption{Panels (a1) and (b1) [or panels (a2) and (b2)] are the same as 
	figure 11(a) [or figure 10] but for different values the ring thickness: 
	$R_{\rm{ring}} = 0.2M$ [(a1) and (a2)] and 0.3$M$ [(b1) and (b2)], respectively.
}\label{fig12}
\end{figure}
\clearpage

\begin{table}
  \tbl{Estimated values of $a$ and $i$ from given $A$ and $C$.}{%
  \scalebox{1.2}[1.2]{
\begin{tabular}{cccc}\hline
quantities & thickness of the ring & spin & inclination angle \\
 $(A,C)$ & $R_{\rm{ring}}/M$ & $a/M$  & $i^{\circ}$ \\ \hline
 $(100,0.9)$ & 0.1& 0.72 & 45 \\
	& 0.2& 0.74 & 46  \\
	& 0.3& 0.76 & 46  \\ \hline
 $(100,1.2)$ & 0.1 & 0.78 & 69 \\
	& 0.2 & 0.77 & 70 \\
	& 0.3 & 0.76 & 69 \\ \hline
 $(300,0.9)$ & 0.1 & -0.44 & 22 \\
	& 0.2 & -0.43 & 22  \\
	& 0.3 & -0.44 & 23 \\ \hline	
 $(300,1.2)$ & 0.1 & -0.27 & 64 \\
	& 0.2 & -0.27 & 64  \\
	& 0.3 & -0.28 & 65  \\ \hline 
    \end{tabular}}}\label{tab3}
\begin{tabnote}
\end{tabnote}
\end{table}

\section{Discussion}
\subsection{Brief summary}			

In this paper, we propose a new method for measuring $a$ and $i$ from observational data 
for given mass $M$.
In this method, we consider a gas cloud, which is assumed to have a ring shape,
falling onto a black hole.
Assuming that the gas ring emits monochromatic radiation,
we calculated the flux variation and the time averaged frequency shift of the radiation
received by a distant observer for various combinations of $a$ and $i$.
We have demonstrated that we can uniquely determine $a$ and $i$ from observational data, 
not critically depending on the thickness of the ring.

The actual procedures of our method are in three steps:
\begin{enumerate}
\item Obtain time-sequence data of normalized flux $F(t)$ and photon numbers $N(t,g)\delta g$
  as functions of frequency or $g$ (frequency-shift factor).
\item Calculate the key quantities, $A$ and $C$, by equations (15)--(17) and (19).
\item Estimate $a$ and $i$ by using figure 10. 
\end{enumerate}
In addition ,
we can also estimate the ring thickness from $A_{1}$ and $A_{2}$ by using figure 11(a), figures 12(a1) and (b1).
Even when the frequency information is not available, we still have a good guess on $a$ solely from $A$, 
especially when the ring thickness is relatively large ($\sim 0.3 M$).

\subsection{Distinctive features of the proposed method}
 
The most distinctive feature of the proposed method is that we can directly measure a black hole spin.
This is made possible by focusing on non-periodic light variation events in
the innermost region at $r<r_{\rm{ms}}$.
By contrast, the widely used methods are based on the steady 
or periodic phenomena occurring at $r\geq r_{\rm{ms}}$. 
In this respect our method is not only independent and complementary to other methods,
but also advantageous in probing the innermost space-time structure around a black hole.

The most popular methods of spin measurements based on continuum or Fe K$\alpha$ line spectra
totally depend on the assumption that the inner edge of the disk is at $r_{\rm{ms}}$ 
(Duro et al. 2011; McClintock et al. 2011; Steiner et al. 2011; McClintock et al. 2014; Reynolds 2014). 
We wish to point that there is no solid physical reason for this assumption;
i.e., the disk could be truncated at a radius greater than $r_{\rm ms}$.
It is of great importance to note the facts
that line and continuum spectra are sensitive more to the radius of the inner edge of 
the emission region than to the spin values. 
This is obvious for the continuum spectra, since the smaller $r_{\rm ms}$ is,
the higher is the maximum disk temperature ($T_{\rm max}$)
and the smaller is the region emitting blackbody radiation with $T_{\rm max}$,
while the shape of the gravitational potential at $r > r_{\rm ms}$
does not largely change with a change in $a$.
This is also true for line spectra. Kojima (1991) compared two line profiles, 
one from a disk with the inner edge at $6M$ around a non-rotating black hole and 
another from a disk with the same radius of the inner edge but around a maximally rotating black hole, 
finding little difference between them (see his figure 2). 
In short, these major methods only indirectly measure spins.

One may think that our method might crucially depend how to take the initial time ($t_{\rm{i}}$) and radius, 
when and where a rotating ring starts to fall. This is not the case, however. 
We can arbitrarily choose the initial radius as long as $r \approx r_{\rm{ms}}$,
as we have argued in subsection 4.1.
This is because the gas ring is initially rotating on nearly a circular orbit 
and so its emitted flux is nearly constant at first, not contributing to $A_1$.
As a result, $A$ does not critically depend on the initial position nor time.

\subsection{Observational implications: shot analysis}
	
Our ring model may possibly be applied to the X-ray shots of black hole binaries. 
X-ray shots are flare-like light variations with sharp peaks (see Oda et al. 1971; \cite{key-11}; Warren et al. 2005) 
and are observed during the so called low/hard state, 
which is characterized by power-law spectra and is associated with substantial variations 
(see e.g., a review by Done et al. 2007).

Negoro et al. (1994) developed the technique of superposed shots,
adding plenty of shot profiles (in X-ray bands) by aligning their peaks (called "superposed shot" analysis: 
see also \cite{key-19}; \cite{key-36}; Yamada et al. 2013).
Furthermore Feng et al. (1999) detected the shot profile with high time resolution, and found that 
(1) the time profile of the superposed shot is rather time symmetric with respect to its peak time
 and that (2) it can well be fit with the sum of two exponential functions:
one with time constants of $\sim 0.01$ [s] and another of $\sim 0.1$ [s]. 

The expected variation timescale by the ring model is
\begin{equation}
 t_{\rm var} \sim 10(M/10M_{\odot})\ \rm ms
\end{equation}
or $\sim 0.01$ s for Cyg X-1 with $12.8\leq M/M_{\odot} \leq 14.8$ (Orosz et al 2011).
This means, 
the light variations predicted by our ring model may be contained within the superposed shot. 
The timescale is more favorably long for supermassive black holes.
Therefore, it might be possible to determine $a$ and $i$ by future observations. 

By means of global MHD simulations 
Machida et al. (2003) studied the time evolution of a torus, which is initially 
threaded by a weak toroidal magnetic field, around a non-rotating black hole.
They carefully analyzed the simulation results and found 
that gas clouds (or streams) with a spiral shape intermittently fall onto the black hole
from the inner edge of the torus. These gas clouds are heated by dissipation of
magnetic energy via magnetic reconnection and are thus expected to emit intense X-rays.
This is exactly the situation that we have postulated in our blob or ring model.

We, however, note that the shape of accreting gas clouds is not a full ring but is a spiral
according to the MHD simulation. It is thus necessary to check whether our proposed model 
based on the full rings can apply to spiral-shaped gas clouds or not,
We calculate a half ring model as the intermediate case between the blob model and the ring model
and show its light curves in figure 13.
The light curves resemble those of the blob model in the sense that they
have several peaks (or humps) due to the gravitational lensing and the beaming,
but the duration of each peak is much longer in the half-ring model.
We also confirm the tendency that the peak flux of the half ring first increases by the focusing effect around $r_{\rm ph}$ and then decays due to gravitational redshift.
Such overall variations are similar to those in the ring model.
We thus expect that the superposed light curves of the half rings (with different initial phases)
are equivalent to those of a full ring (recall figure 6).
In conclusion, we can use the ring model for spin measurements regardless of the precise shape of 
the gas cloud; it can be a blob, a half ring, or a full ring.

\begin{figure}
\begin{center}
\includegraphics[width=12.0cm]{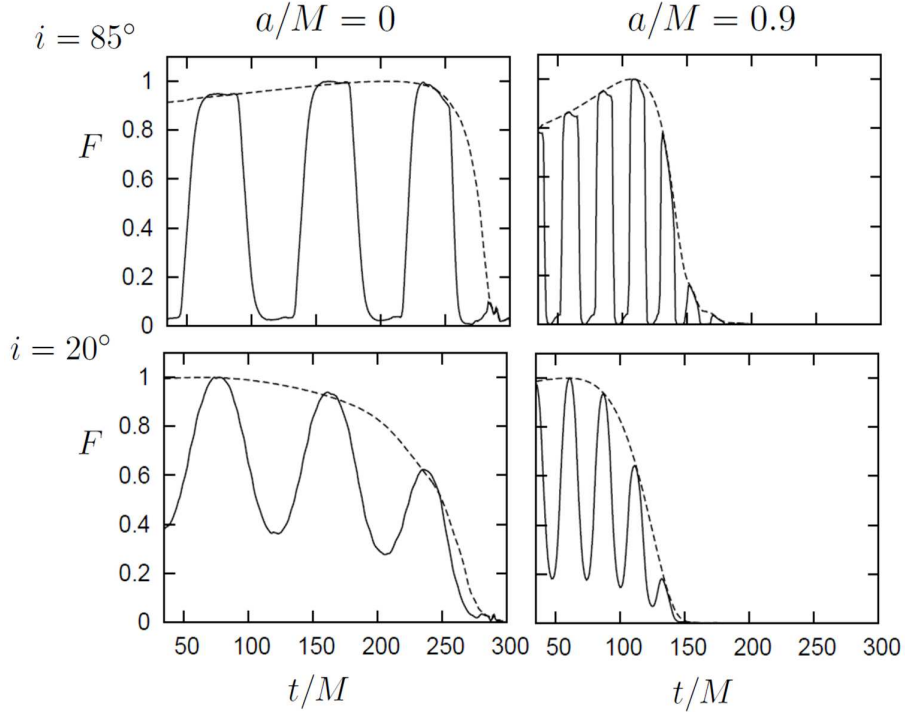}
\end{center}
\caption{Light curves of a half ring (by the solid line) 
and those of a full ring (by the dashed line) 
for $a/M=0$ (left panels) and 0.9 (right panels), and for $i$=85$^\circ$ (upper) and 
20$^\circ$ (lower), respectively. We assumed $R_{\rm ring} = 0.1M$ in all models.
}\label{fig13}
\end{figure}

\subsection{Observational implications: feasibility}

How accurately can we measure the spin value when applying our method to the actual X-ray observational data? 
In subsection 4.3, we have demonstrated that we can estimate the spin value by measuring $A$. 
In this subsection, we evaluate an accuracy of the spin measurement by performing Monte Carlo simulation for light curve of the gas ring.

Let us first examine the case of observing Cyg X-1.
The time resolution of the RXTE is 1~[ms] 
(Feng et al. 1999; Liu $\&$ Li 2004; Wu et al. 2007).
This timescale corresponds to $0.07t_{\rm var}$ for the black hole mass of $M=14M_{\odot}$ 
(Orosz et al 2011), 
where, $t_{\rm var}$ is the time scale of flux variation [see equation (26)].
Hereafter we thus assign a constant time bin; $\Delta t=0.07t_{\rm var}$. 
Suppose that we observe Cyg X-1 with the RXTE satellite.
According to Feng et al. (1999), however, the number of photons that are received 
during the time bin of $0.07 t_{\rm var}$ is only 2 [counts/keV] even at the peak.
We thus give up using the RXTE data and calculate the cases of observations
with future satellites, such as Athena (or LOFT). 
Since the effective area of these missions is about 10 (100) times 
as large as that of the RXTE, the average count rate around the peak
is $N_{\rm max}=20$ (200) [counts/(0.07$t_{\rm var}$)/keV].

The actual procedures of evaluating the spin parameter ranges ($a_{\rm est}$)
are as follows:

\begin{enumerate}
\item
We first choose one value of the spin parameter ($a_{\rm true}$) and, 
assuming that the photon number at the flux maximum to be $N_{\rm max}$ given above,
we calculate the expected (averaged) number of photons [$N(t)$] received per each time bin 
with a length of  $0.07 t_{\rm var}$, noting $N(t)$ $\propto F(t)$. 
\item
On the averaged photon number profile [$N(t)$] we superpose random errors [$\delta N_i(t)$] 
(with $i=1, 2, \cdots$), assuming that photon statistics obeys Poisson distribution 
[with the standard deviation of $\sigma_N= \sqrt{N(t)}$]. 
Figure 14 give one such example. 
\item
We superpose 500 such shot profiles, 
$N_{\rm tot} (t)\equiv \sum_i [N(t)+\delta N_i(t)]$. 
\item
We then normalize the time sequence of the number counts [$N_{\rm tot}(t)$] of each run 
by their maximum value, and call it the re-defined flux distribution, $\tilde{F}(t)$. 
\begin{figure}
\begin{center}
\includegraphics[width=8.0cm]{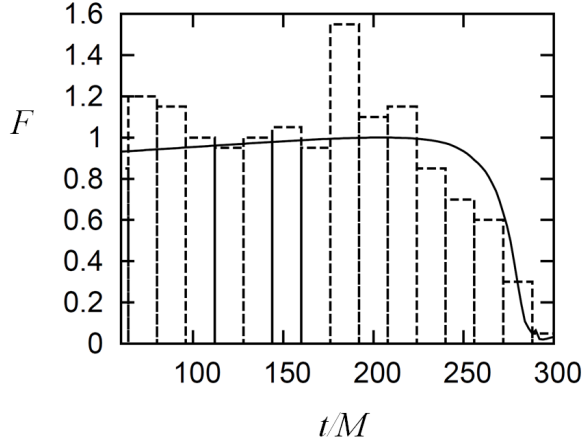}
\end{center}
\caption{
An example of the time profile of the observational flux, in which
random errors are superposed by assuming the Poisson statistics. 
The solid curve represents $F(t)$, the result of figure 7,
while the vertical bars represent one example of $N(t)+\delta N_{i}(t)$. 
Here, we postulate the case that we observe Cyg X-1 with Athena,
and the model parameters are $(a_{\rm true}/M,i)=(0,85^{\circ})$.
The time interval corresponds to $0.07 t_{\rm var}\simeq 14M$.
}\label{fig14}
\end{figure}
\item
We calculate $A$ by inserting $\tilde{F}(t)$ to $F(t)$ in equations (15)--(17) 
and estimate the spin parameter ($a_{\rm est}$) by using figure 10.  
\item
We repeat the same process for $10^{4}$ times, plot the distribution of the estimated spin, 
$a_{\rm est}$, and evaluate its 1-$\sigma$ confidence range. 
\end{enumerate}

We summarize the results in table 4 for each of observations with Athena and LOFT, respectively. 
In the case of Athena (or LOFT), we can determine the spin value with following error, $\delta a/M=  0.06-0.39$ ($0.03-0.14$).

\begin{table}
  \tbl{estimated spin values of Cyg X-1.}{
  \scalebox{1.2}[1.2]{
\begin{tabular}{cccccccc}\hline
 & Athena & & LOFT & &\\ 
 $a_{\rm{true}}/M$  & $a_{\rm est}$ for $i=85[^\circ])$ & $i=40[^\circ])$ & $i=85[^\circ])$ & $i=40[^\circ])$  \\ \hline
0.8   &	0.77--0.83			&	0.68--0.85				&	0.78-- 0.81		&	0.79--0.85				\\	
0.4   &	0.31--0.46			&	0.28--0.50					&	0.35--0.40			&	0.38--0.50				\\
0     & 	-0.15--0.11			&	-0.15--0.10			&	-0.10--0.02			&	 -0.01--0.09			\\
-0.4 &	-0.54-- -0.22		&	-0.51-- -0.31					&	-0.48-- -0.35		&	-0.49-- -0.35		\\
-0.8 &	-1.0-- -0.61		&	-1.0 -0.70					&	-0.87-- -0.74		&	-0.88-- -0.74		\\ \hline
\end{tabular}}}\label{tab4}
\begin{tabnote}
\end{tabnote}
\end{table}

Similar analysis of the case of Cyg X-1, we estimate the accuracy of spin in the case of MCG-6-30-15. 
The time resolution of RXTE is $10^{2}$[s]$=0.1t_{\rm var}$, where we postulate the mass is 
$10^{6}M_{\odot}$ (Lee et al. 2000; Reynolds 2000).
Then we set to $\Delta t=0.1t_{\rm var}$. 
The averaged photon count rate of one shot peak is 100 [counts/(0.1$t_{\rm var}$)/keV]. 
Further, we set the number of superposed shots to $10$[shot], since the time interval between start and end time of the observation, 
$10^{5}$ [s] is about 10 times as large as an time duration of the shots (Lee et al. 2000; Reynolds 2000). 
In addition, since the effective area of Athena (or LOFT) is about 10 (100) times 
as large as that of the RXTE, the average count rate around the peak
is $N_{\rm max}=10^{3}$ ($10^{4}$) [counts/(0.1$t_{\rm var}$)/keV].

The results for MCG-6-30-15 are summarized in table 5. 
By using Athena (or LOFT), we can estimate spin value with following error, $\delta a/M= 0.07-0.45$ ($\delta a/M= 0.03-0.14$), 
while in the case of RXTE, we can scarcely detect the spin value with the rough accuracy, $\delta a/M =0.13-0.78$.

\begin{table}
  \tbl{estimated spin values of MCG-6-30-15.}{
  \scalebox{1.2}[1.2]{
\begin{tabular}{cccccccc}\hline
 & RXTE && Athena & & LOFT & &\\ 
 $a_{\rm{true}}/M$  & $a_{\rm est}$ for $i=85[^\circ])$ & $i=40[^\circ])$ & $i=85[^\circ])$ & $i=40[^\circ])$ & $i=85[^\circ])$ & $i=40[^\circ])$ \\ \hline
0.8   & 	0.74--0.87		&	0.77--0.83			&	0.68--0.85			&	0.73--0.80				&	0.78-- 0.81		&	0.79--0.85				\\	
0.4   & 	0.21--0.58 	&	0.32--0.46		&	0.28--0.50			&	0.34--0.49					&	0.35--0.40			&	0.38--0.50				\\
0     & 	 -0.30--0.33		&	-0.13--0.10	    & 	-0.16--0.10			&	-0.15--0.13			&	-0.1--0.02			&	 -0.01--0.09			\\
-0.4 & 	-0.70--0.08		&	-0.53-- -0.16		&	-0.55-- -0.22		&	-0.51-- -0.31					&	-0.48-- -0.34		&	-0.49-- -0.35		\\
-0.8 & 	-1.0-- -0.25	   &	-1 -- -0.56	&	-1.0-- -0.55		&	-1.0-- -0.70					&	-0.88-- -0.74		&	-0.88-- -0.74		\\ \hline
\end{tabular}}}\label{tab4}
\begin{tabnote}
\end{tabnote}
\end{table}

\subsection{Remaining issues}
\
We have so far argued that the proposed method should in principle work, 
however we recognize number of essential issues to be considered before 
applying this method to actual observational data.
We first evaluate what resolutions are necessary for spin determination.

The first issue is time resolution.
As we indicated by equation (26), the time resolution required by the ring model
is less than $10 (M/10M_{\odot })$ ms.
This value itself is attainable but we need good photon statistics, which is another concern.
Note that this timescale is much longer than the dynamical timescale (or the period of orbital motion),
which is on the order of less than 1 ms. This is because we are considering gas 
falling with keeping a finite angular momentum.

The next one is energy resolution.
In order to estimate $i$ from the $C$-value within an error of $20^\circ$, 
the required energy resolution is $0.1 \times E_{\rm line}$ from figure 10, 
where $E_{\rm{line}}$ is the energy of the fluorescent Fe K$\alpha$ line 
in the inertial frames of the emitting gas.

Finally, we need to know $M$ with high accuracy to estimate $a$ accurately.
As we mentioned in subsection 4.2, we can measure $a$ mainly through $A$, and
$A$ depends on $M$ [see equations (15)-(17)]. 
Let us denote $\delta M$ as an error in $M$ estimation. Then the error in $A$ is evaluated as
\begin{equation}
  \frac{\delta A}{A} = \frac{\delta M}{M}\left( 1+\frac{\delta M}{M}\right)^{-1}.
\end{equation} 
Using this relation we calculate the range of the estimated spin, $a_{\rm{est}}$,
for a given (true) spin value, $a_{\rm{true}}$, for two cases with $\delta M/M = 0.1$ and $0.2$.
The results are listed in table 6.
We see in this table that we can determine $a/M$ with good accuracy of within $\sim$ 0.05
for high spin cases, or less than 0.20 for low spin cases (say, $a\sim 0$), 
if the mass determination accuracy is $\delta M/M = 0.1$.
When $\delta M/M = 0.2$ the situation gets worse, but
the spin determination accuracy is still $\sim$ 0.3 even for low spin cases.

\begin{table}
  \tbl{relationship between the estimated spin and true one for a given error in mass.}{%
  \scalebox{1.2}[1.2]{
\begin{tabular}{ ccc}\hline
  mass error & true spin & estimated spin \\ 
  $|\delta M|/M$ & $a_{\rm{true}}/M$ & $a_{\rm{est}}/M$ \\ \hline
	0.1 & 0.8   & 0.74 -- 0.85 \\ 
     & 0.4   & 0.30 -- 0.49  \\
     & 0     &  -0.17 -- 0.12\\ 
     & -0.4 & -0.55 -- -0.23 \\ 
     & -0.8 & -1.0 -- -0.62 \\ \hline
	0.2 & 0.8   & 0.69 -- 0.92 \\ 
     & 0.4   & 0.21 -- 0.60  \\
     & 0     & -0.31  -- 0.28\\ 
     & -0.4 & -0.68 -- -0.021 \\
     & -0.8 & -1.0 -- -0.40\\ \hline
    \end{tabular}}}\label{tab6}
\begin{tabnote}
\end{tabnote}
\end{table}

Finally, we give a summary of future issues.
\begin{enumerate}
\item We considered particular shape of infalling gas cloud and 
    assumed no changes of the shape with time.
   That is, a distortion of the ring shape in a deep potential well was not considered,
   We need to relax this assumption and improve the method by using, e.g., MHD simulation data.
\item  The temperature and the emissivity ($j_0$) of the ring is kept constant.
    This may not be so realistic, since heating by magnetic energy release and/or radiative
    cooling should be effective.
\item We assumed monochromatic radiation form the gas clouds for the sake of simplicity.
    This is a critical assumption, on which the proposed method is constructed, but its physical
    reason is not yet clear. We need to expand our methodology so as to incorporate more general
    cases of continuum radiation etc.
\item Actual correspondence with real observational data is under consideration.
      It will be important to identify which data really represent an infall of a gas ring
      in connection with real observational data and MHD simulation data.
\end{enumerate}



\begin{thebibliography}{}
\bibitem[Bardeen et al. (1972)]                         {key-1} Bardeen, J. M., Press, W. H.,\& Teukolsky, S. A.\ 1972, \apj, 178, 347 
\bibitem[Blum et al. (2009)]                              {key-2} Blum, J. L., et al.\ 2009, \apj, 706, 60 
\bibitem[Bradt et al. (1993)]                             {key39} Bradt, H. V., Rothschild, R. E., Swank, J. H. \ 1993 \aap, 97, 355
\bibitem[Carter (1968)]                                    {key-3} Carter, B.\ 1968, Phys. Rev., 174, 1559
\bibitem[Cunningham $\&$ Bardeen (1973)]                            {key-4} Cunningham, C. T.,\& Bardeen, J. M. 1973, \apj, 183, 237
\bibitem[Done et al. (2007)]                             {key-5} Done, C., $\rm{Gierli\acute{n}ski}$, M.,\& Kubota, A.\ 2007, \aapr, 15, 1
\bibitem[$\rm{Dov\check{c}iak}$ et al. (2004a)]  {key-6} $\rm{Dov\check{c}iak}$, M., Bianchi, S., Guainazzi, M., Karas, V.,\& Matt, G.\ 2004a, \mnras, 350, 745
\bibitem[$\rm{Dov\check{c}iak}$ et al (2004b)] {key-7} $\rm{Dov\check{c}iak}$, M., Karas, V.,\& Yaqoob, T.\ 2004b, \apjs, 153, 205
\bibitem[Duro et al (2011)]                                {key-8} Duro, R., et al.\ 2011, \aap, 533, L3 
\bibitem[Feng et al. (1999)]								{key-9} Feng, Y. X., Li, T. P., \& Chen, L.\ 1999, \apj, 514, 373 
\bibitem[Ghez et al. (2005)]                              {key-10} Ghez, A. M., et al.\ 2005, \apj, 620, 744
\bibitem[$\rm{Gierli\check{n}ski}$ $\&$ Zdziarski (2003)] {key-11} $\rm{Gierli\check{n}ski}$, M., \& Zdziarski, A.A. 2003, \mnras, 343, L84
\bibitem[Hanawa (1989)]                                   {key-12} Hanawa, T. 1989,\ \apj, 341, 948
\bibitem[Karas et al. (1992)]                              {key-13} Karas, V., $\rm{Vokrouhlick\acute{y}}$, D.,\& Polnarev, A. G.\ 1992, \mnras, 259, 569
\bibitem[Kato et al. (2008)]                              {key-14} Kato, S., Fukue, J.,\& Mineshige, S.\ 2008, Black-Hole Accretion Disks -- Towards a New Paradigm (Kyoto: Kyoto University Press)
\bibitem[Kato (2001)]                                       {key-15} Kato, S.\ 2001, \pasj, 53, 1
\bibitem[Kojima (1991)]                                    {key-16} Kojima, Y.\ 1991, \mnras, 250, 629	
\bibitem[Laor (1991)]                                       {key-17} Laor, A. \ 1991, \apj, 376,90
\bibitem[Lee et al. (2000)]                                {key-40} Lee, J. C., Fabian, A. C., Reynolds, C. S., Brandt, W. N., Iwasawa, K. \ 2000, \mnras, 318, 857
\bibitem[Li et al. (2005)]                                   {key-18} Li, L.-X., Zimmerman, E. R., Narayan, R.,\& McClintock J. E.\ 2005, \apjs, 157, 335
\bibitem[Liu $\&$ Li (2004)]								{key-19} Liu, C. Z. \& Li, T. P.\ 2004, \apj, 611, 1084 
\bibitem[Machida $\&$ Matsumoto (2011)]         {key-20} Machida, M., \& Matsumoto, R., \ 2003, \apj, 585, 429
\bibitem[McClintock et al. (2011)]                     {key-21} McClintock, J. E et al.\ 2011, Class. Quantum Grav., 28, 114009
\bibitem[McClintock et al. (2014)]                     {key-22} McClintock, J. E., Narayan, R.,\& Steiner, J. F.\ 2014, \ssr, 183, 295
\bibitem[Negoro et al. (1994)]                           {key-23} Negoro, H., Kitamoto, S., Takeuchi, M.,\& Mineshige, S.\ 1994, \apj, 423, L127
\bibitem[Oda et al. (1971)]                                {key-24} Oda, M., Gorenstein, P., Gursky, H., Kellogg, E., Schreier, E., Tananbaum, H.,\& Giacconi, R.\ 1971, \apj, 166, L1 
\bibitem[Orosz et al. (2011)]                             {key-25} Orosz, J. A., McClintock, J. E., Aufdenberg, J. P., Remillard, R. A., Reid, M. J., Narayan, R.,\& Gou, L.\ 2011, \apj, 742, 84
\bibitem[Rezzolla et al. (2003)]                         {key-26} Rezzolla, L., Yoshida, S'i., Maccarone, T. J.,\& Zanotti, O.\ 2003, \mnras, 344, L37
\bibitem[Remillard (2005)]                                {key-27} Remillard, R. A.\ 2005, AN, 326, 804
\bibitem[Reynolds (2000)]                                {key-41} Reynolds, C. S.,\ 2000, \apj, 533,820
\bibitem[Reynolds (2014)]                                {key-28} Reynolds, C. S.,\ 2014, \ssr, 183,277
\bibitem[Shahbaz et al. (1999)]                         {key-29} Shahbaz, T., van der Hooft, F., Casares, J., Charles, P. A.,\& van Paradijs J.\ 1999, \mnras, 306, 89
\bibitem[Shapiro et al. (1983)]                          {key-30} Shapiro, S. L.,\& Teukolsky, S. A.\ 1983, Black Holes, White Dwarfs, and Neutron Stars, (New York: John Wiley \& Sons)
\bibitem[Shafee et al. (2006)]                            {key-31} Shafee, R., McClintock, J. E., Narayan, R., Davis, S. W., Li, L.-X., \& Remillard, R. A.\ 2006, \apj, 636, L113
\bibitem[Shidatsu et al. (2014)]                        {key-32} Shidatsu, M., et al. 2014, \apj, 789, 100 
\bibitem[Steiner et al. (2011)]                          {key-33} Steiner, J. F., et al.\ 2011, \mnras, 416, 941
\bibitem[Strohmayer (2001)]                            {key-34} Strohmayer, T. E.\ 2001, \apj, 552, L49
\bibitem[Tanaka et aｌ. (1995)]                           {key-35} Tanaka, Y., et al.\ 1995, \nat, 375, 659
\bibitem[Wu et al. (2007)]								  {key-36} Wu, Y.-X., Liu, C.-Z., \& Li, T.-P.\ 2007, \apj, 660, 1386 
\bibitem[Yamada et al. (2013)]                          {key-37} Yamada, S., Negoro, H., Torii, S., Noda, H., Mineshige, S., \& Makishima, K.\ 2013, \aplett, 767, L34
\bibitem[Yoshida (1993)]                                  {key-38} Yoshida, H., \ 1993, CeMDA, 56, 27Y 
\end{thebibliography}
\end{document}